\documentclass[conference]{IEEEtran}
\IEEEoverridecommandlockouts
\usepackage{cite}
\usepackage{amsmath,amssymb,amsfonts}
\usepackage{algorithmic}
\usepackage{graphicx}
\usepackage{textcomp}
\usepackage{xcolor}
\usepackage{hyperref}
\usepackage{graphicx}
\usepackage{url}
\usepackage{color}
\usepackage{syntax}
\usepackage{newfloat}
\usepackage{multicol}
\usepackage{tikz}
\usepackage{filecontents} 
\usepackage{graphicx}
\usepackage{epstopdf} 
\usepackage{xspace}
\usepackage{enumitem}
\usepackage{svg}
\usepackage[ruled,linesnumbered,noend]{algorithm2e}
\SetKw{Continue}{continue}
\usepackage{adjustbox}
\usepackage{wrapfig}
\usepackage{graphbox}
\usepackage{rotating}
\usepackage{pdfpages}
\usepackage{mdwlist}
\usepackage{multirow}
\usepackage{booktabs}
\usepackage{makecell}

\usepackage{subcaption}
\usepackage[labelfont=bf,skip=0pt]{caption}

\usepackage{hyperref} 
\hypersetup{
	citebordercolor={0 1 0},
	citecolor=blue,
	colorlinks=false,
	filebordercolor={0 .5 .5},
	filecolor=green,
	linkbordercolor={1 0 0},
	linkcolor=blue,
	menubordercolor={1 0 0},
	pageanchor=true,
	pagebackref=true,
	pdfborder={0 0 1},
	pdfpagelabels=true,
	pdftex,
	plainpages=false,
	urlbordercolor={0 1 1},
	urlcolor=blue,
}

\usepackage[capitalize,nameinlink]{cleveref}
\hypersetup{%
	bookmarksnumbered, bookmarksopen=true, bookmarksopenlevel=1,%
}
\crefname{figure}{Figure}{Figures}
\crefname{listing}{Query}{Queries}
\crefname{section}{Section}{Sections}
\crefname{table}{Table}{Tables}
\crefname{BNF}{Grammar}{Grammars}
\crefname{algorithm}{Algorithm}{Algorithms}
\crefname{equation}{Equation}{Equations}

\usepackage{listings}
\definecolor{mygreen}{rgb}{0,0.6,0}
\definecolor{mygray}{rgb}{0.5,0.5,0.5}

\newcommand{\incode}[1]{\lstinline{#1}}

\DeclareFloatingEnvironment[
fileext   = logr,
listname  = {List of Grammars},
name      = Grammar,
placement = !htbp
]{BNF}

\usepackage{mdframed}

\newcommand{\msim}{\raise.17ex\hbox{$\scriptstyle\sim$}}

\newcommand{\ra}[1]{\renewcommand{\arraystretch}{#1}}
\newcommand{\myparatight}[1]{\smallskip\noindent{\bf {#1}.}}

\newcommand{\eat}[1]{}
\newcommand{\eg}{e.g.,\xspace}
\newcommand{\ie}{i.e.,\xspace}
\newcommand{\aka}{a.k.a,\xspace}

\newcommand{\tool}{\textsc{ProGQL}\xspace}
\newcommand{\depimpact}{\textsc{DepImpact}\xspace}
\newcommand{\pa}{\textsf{PA}\xspace}
\newcommand{\pg}{\textsf{PG}\xspace}
\newcommand{\pgs}{\textsf{PG}s\xspace}

\usepackage{pifont}

\newcommand{\distance}{6pt}
\def\BibTeX{{\rm B\kern-.05em{\sc i\kern-.025em b}\kern-.08em
    T\kern-.1667em\lower.7ex\hbox{E}\kern-.125emX}}

\newcommand{\rev}[1]{{#1}}

\begin{document}

\title{\tool: A Provenance Graph Query System for Cyber Attack Investigation \\
\vspace*{1ex}\LARGE Technical Report}


\author{
 \IEEEauthorblockN{Fei Shao\IEEEauthorrefmark{2}, Jia Zou\IEEEauthorrefmark{1}, Zhichao Cao\IEEEauthorrefmark{1}, Xusheng Xiao\IEEEauthorrefmark{1}}
    \IEEEauthorblockA{\IEEEauthorrefmark{1}Arizona State University.
    Email: \{jia.zou, zhichao.cao, xusheng.xiao\}@asu.edu}
    \IEEEauthorblockA{\IEEEauthorrefmark{2}Case Western Reserve University.
    Email: fxs128@case.edu}
}


\maketitle

\begin{abstract}

Provenance analysis (\pa) has recently emerged as an important solution for cyber attack investigation.
\pa leverages system monitoring to monitor system activities as a series of system audit events and organizes these events as a \textit{provenance graph} to show the dependencies among system activities, which can reveal steps of cyber attacks. 
\rev{Despite their potential, existing \pa techniques face two critical challenges: (1) they are inflexible and non-extensible, making it difficult to incorporate analyst expertise, and (2) they are memory-inefficient, often requiring $>$100GB of RAM to hold entire event streams, which fundamentally limits scalability and deployment in real-world environments.}
\rev{To address these fundamental limitations, we propose the \tool framework, which provides a domain-specific graph search language with a well-engineered query engine,
allowing \pa over system audit events and expert knowledge to be jointly expressed as a graph search query and thereby facilitating the investigation of complex cyberattacks.
In particular, to support dependency searches from a starting edge required in \pa, \tool introduces new language constructs for constrained graph traversal, edge weight computation, value propagation along weighted edges, and graph merging to integrate multiple searches. 
Moreover, the \tool query engine is optimized for efficient incremental graph search across heterogeneous database backends, eliminating the need for full in-memory materialization and reducing memory overhead.}
Our evaluations on real attacks demonstrate the effectiveness of the \tool language in expressing a diverse set of complex attacks compared with the state-of-the-art (SOTA) graph query language Cypher, 
and the comparison with the SOTA \pa technique \depimpact further demonstrates the significant improvement of the scalability brought by our \tool framework's design ($8\times$ saving of memory consumption without penalty on runtime performance).
\end{abstract}


\UseRawInputEncoding
\section{Introduction}
\label{sec:intro}
Large enterprises are increasingly plagued by cyber-attacks, causing significant financial losses~\cite{ebay,opm,target,homedepot,ya:yahooleak,equifax,marriott}.
These attacks often exploit multiple types of vulnerability to infiltrate target systems in multiple stages, posing challenges for detection and investigation.
To counter these attacks, recent approaches based on \emph{ubiquitous system monitoring} have emerged as an important approach that monitors system activities and assists attack investigation~\cite{backtracking,backtracking2,wormlog,logtracking,mcitracking,liu2018priotracker,gao2018aiql,gao2018saql}.
System monitoring collects kernel auditing events about system calls as system audit logs.
The collected data enables techniques based on causality analysis to perform \textit{provenance analysis} (referred to as \pa) on attack-related events ~\cite{backtracking,backtracking2,liu2018priotracker,mcitracking,ma2016protracer}, which provides the contextual information about the attacks to identify entry points of invasions (i.e., backward \pa) and ramifications of attacks (i.e., forward \pa).

\pa assumes causal dependencies between system entities (\eg files, processes, and network connections) involved in the same system call event (\eg a process reading a file).
Based on this assumption, given a Point-Of-Interest (POI) event (\eg an alert reported by intrusion detection),
these techniques search for the system call events that have dependencies on the POI event and organize these events as a \textit{provenance graph} (referred to as \pg), a data structure that models dependencies between system activities~\cite{hassan2019nodoze,backtracking,backtracking2}, with nodes representing system entities (\eg processes and files) and edges representing events (\eg file creation by a process).
A \pg can provide contextual information for the POI event by reconstructing a chain of events that lead to the POI event. 
It has been proven effective in reducing false alerts of intrusions~\cite{alertfp2,ransomware} and assisting timely system recoveries~\cite{taser,intrusionrecovery}.
For example, as ransomware and compression programs (\eg \incode{bzip2}) both read and write many files in a short period of time, many ransomware detectors that check only the behavior of a single process will likely classify \incode{bzip2} as ransomware;
with the dependency graph provided by causality analysis, the detector can distinguish \incode{bzip2} from ransomware, as the entry point of ransomware (\eg email attachment) is often different from the \incode{bzip2} program.

While \pa makes promising progress in attack investigation, existing techniques~\cite{backtracking,backtracking2,taser,intrusionrecovery,liu2018priotracker} suffer from \underline{two fundamental limitations}.
First, existing techniques \textbf{lack \textit{extensibility} to incorporate expert knowledge and \textit{flexibility} to integrate multiple \pa results.}  
\rev{\pa is prone to dependency explosion since existing techniques mainly use happen-before relationships to identify dependencies~\cite{reduction,backtracking}. Consequently, the resulting \pg often exceeds one million edges. 
To mitigate this, recent techniques define edge weights based on edge attributes and apply optimization techniques to filter irrelevant dependencies\cite{taser,intrusionrecovery,mcitracking,liu2018priotracker,ma2016protracer,depcomm}.
However, these techniques are not universally effective across all attack scenarios, and investigations often require applying multiple \pa across several alerts~\cite{depimpact} to determine which are related and which are irrelevant.
This process is similar to interactive queries in database systems, where results are progressively refined to support data-driven decisions. To facilitate this process, expert knowledge must be incorporated into provenance analysis for choosing weight functions or specifying domain-specific filters, and the results from multiple analyses need to be integrated seamlessly. Unfortunately, existing techniques provide neither effective expert knowledge incorporation nor efficient integration of results.}
Second, \textbf{existing techniques are memory-hungry and inefficient.} For example, they often load all the events within a period of time into memory and perform analysis on these events to filter irrelevant ones. 
This strategy requires a significant amount of memory to hold all the events (often $>100GB$)~\cite{depimpact,HOLMES,hassan2019nodoze} and greatly limit the scalability of \pa and prevent it from being used in resource-constrained environments. 

\noindent\textbf{Contribution}.
\rev{To address these fundamental limitations of existing \pa techniques, in this paper, we propose a novel query framework, called \tool, which provides (1) a domain-specific graph search language that can jointly express \pa over system audit events and expert knowledge as a graph search query, and (2) a query engine that optimizes query execution by combining domain-specific optimizations with database backends.} 
The language design of \tool is motivated by the following insights. 
\rev{
\begin{itemize}[noitemsep, topsep=1pt, partopsep=1pt, listparindent=\parindent, leftmargin=*]
\item \textit{Dependency-based Graph Search}: Fundamentally, \pa is a graph search technique, finding edges that act as dependencies of a specific edge (\ie POI event) in a \pg derived from system audit logs.
For example, backward \pa finds the edges whose start time is before the end time of the POI event, and then recursively applies the same search constraints on the last found edges until no more edges can be found.  
Also, computing certain edge properties requires \textit{value propagation}, such as finding all the ancestors for an edge.
While existing graph search languages (e.g., Cypher~\cite{cypher}, SPARQL~\cite{sparql,sparcle05}, and CRPQ~\cite{graphquery,graphquery2}) and domain-specific languages (e.g., AIQL~\cite{aiql} and SAQL~\cite{saql}) support finding subgraphs or paths expressible through regular expressions and anomaly patterns, they lack the ability to model dependency-based searches that involve iteratively extending from existing edges and checking edge-dependent constraints. 
For example, Cypher can match edges such as $e(u,v)$ using attribute-based conditions (e.g., \incode{MATCH (c:node {name: “Chrome”})}), but it cannot express queries that depend on the properties of adjacent edges. 
Specifically, it cannot retrieve incoming edges of $u$ whose start time is earlier than the maximum end time among $v$’s outgoing edges.
\item \textit{Edge Weights and Value Propagation}: Edge weights are used by various \pa techniques to filter out irrelevant dependencies~\cite{depimpact,hassan2019nodoze,liu2018priotracker}, and impact scores propagated from the POI event through the weighted edges are also used to identify attack-related entry nodes~\cite{depimpact,tagpropagation,sleuth}.
Unfortunately, existing languages also lack the capabilities to express edge weights and impact scores computed based on value propagation through weighted edges. 
Additionally, they do not support combining the results of \pa applied to multiple POI events, a capability essential for retaining attack-related information while filtering irrelevant edges. 
\end{itemize}
}

\rev{
To address these issues, we propose a novel graph query language, \tool, built upon the syntax of Cypher~\cite{cypher}, which is a popular graph query language for Neo4j~\cite{neo4j}.
\tool provides novel language constructs for (1) recursive constrained graph search, (2) edge weight assignment based on arbitrary feature projections, and (3) impact score propagation across the graph, and graph merging:
\begin{itemize}[noitemsep, topsep=1pt, partopsep=1pt, listparindent=\parindent, leftmargin=*]
\item \textit{Constrained Graph Search}: \tool extends the \incode{BFS} and \incode{DFS} constructs to support Breadth-First and Depth-First Search with user-defined constraints, allowing dependency edges to be identified based on the properties of adjacent edges.
\item \textit{Edge Weight Computation}: \tool extends the \incode{UNWIND} and \incode{SET} operators of Cypher to support the edge weight computation based on adjacent edges' properties such as time. 
\item \textit{Value Propagation}: \tool extends the Cypher \incode{MATCH} and \incode{SET} operators to support the value propagation through the weighted edges.
\item \textit{Graph Merge}: \tool provides the language constructs (\incode{UNION} and \incode{INTERSECT}) to merge multiple \pgs by combining all edges or retaining only the edges common to each graph. 
\end{itemize}
}

To support the query execution of \tool language, our framework provides a data importer that parses the events recorded in system auditing logs, builds the data model of the parsed events, and performs batch insertion of the modeled events into a database backend with high efficiency.
In particular, the database backend can be built using different types of databases such as relational databases (\eg PostgreSQL~\cite{postgresql} and MyRocks~\cite{myrocks}) and graph databases  (\eg Neo4j~\cite{neo4j} and Nebula~\cite{nebula}). 
Different types of databases support security-related searches with varying efficiency, each excelling at different search types, as shown in recent studies~\cite{aiql,aiqldemo,saql,saqldemo}.
For example, graph databases can efficiently support finding neighboring edges given a POI event, while relational databases can leverage join to efficiently find out events satisfying specific constraints even when the edges are not connected.
Building upon a database backend enables our \tool framework to leverage the critical services provided by the mature infrastructures, such as data management, indexing mechanisms, recovery, and security.

In addition, our framework provides a novel query engine that efficiently executes \tool queries. As \pa usually needs to process a colossal amount of log data~\cite{reduction,reduction2,loggc,depimpact,depcomm,reduction3,reduction4}, 
it is critical to optimize the performance of graph search when building \pgs from system auditing logs. 
\rev{Thus, our query engine optimizes the proposed graph search and merging through an incremental graph search by performing fine-grained edge fetch with the help of the database backend rather than loading the whole graph into memory, greatly improving memory efficiency and search performance.
Moreover, the query engine performs edge merge based on time differences to minimize duplicate information, performs edge weight assignment based on arbitrary feature projections, and supports efficient graph merge through edge signatures. }

\noindent\textbf{Evaluation}. 
We conduct comprehensive evaluations of \tool by composing \tool queries for 14 real attacks that represent a diversified set of attack scenarios. 
The number of the system events to search for each attack case is about 19 million on average.
The results show that \tool can execute these queries using averagely 5GB memory and finish the search within 224 seconds, and the output \pgs preserve all the attack steps for each attack.
To demonstrate the expressiveness of \tool, we compare \tool with Cypher in expressing attack behaviors, and the results show that due to the lack of expressiveness in constrained graph search, weight computation, and value propagation, Cypher needs to use 32 separate queries to achieve the same search results as a \tool query.
\rev{On average, when expressing queries for the same attacks, \tool uses $9\times$ fewer constraints, $15\times$ fewer words, and $17\times$ fewer characters than Cypher.}
Furthermore, \tool queries are executed 22.76 times more efficiently (21s v.s. 478s) and requires only about half ($59.8\%$) of the memory used by Cypher running in Neo4j.

We also compared \tool with the state-of-the-art (SOTA) \pa: \depimpact~\cite{depimpact}. 
The results show that \tool can achieve similar runtime performance as \depimpact with $8\times$ smaller of memory consumption. 
This is a great improvement on scalability, as \depimpact loads all events into memory and performs the in-memory graph search, while \tool leverages the database backends.
Even though the queries with the database may cause performance overhead compared to executing the queries in memory, \tool's edge indexes and the incremental graph search employed by \tool speed up the subsequent search with substantially lower memory consumption, while achieving similar runtime performance. 
We also showed that \tool can easily express \pa techniques with different weight computations and compare their performance.
Finally, we compared the performance of using different types of database backends, the results show that overall Neo4j achieves the best runtime performance and PostgreSQL achieves the lowest memory consumption.
Our code and data are publicly available at our project website~\cite{progql}.

\section{Background and Motivation}

\begin{table}[t]
\caption{Representative system calls}
\ra{1.2}
\centering
\begin{adjustbox}{width=\linewidth}

\begin{tabular}{ll}
\hline
 \multicolumn{1}{c}{\textbf{Event Category}}&  \multicolumn{1}{c}{\textbf{System Call}}\\ \hline
File event & read, write, execute, rename  \\
Process event & execve, execute, clone  \\
Network event & read, write, sendto, sendmsg, recvfrom \\ \hline
\end{tabular}
\end{adjustbox}
\label{tab:events}
\end{table}


\subsection{System Audit Logs}
\label{subsec:system-monitoring}
System audit logs record the kernel-level audit events about system calls and are crucial for cyber attack investigation~\cite{backtracking,backtracking2,wormlog,logtracking,liu2018priotracker,gao2018aiql,gao2018saql,mcitracking}.
These audit events provide detailed information on monitored system calls, describing how system entities interact with system resources and other system entities in a monitored computer system. 
Formally, a system audit event is modeled as a directed edge between subject and object or vice-versa, represented as $\langle$sub, op, obj$\rangle$ or $\langle$obj, op, sub$\rangle$, 
where $sub$ represents a process entity, $obj$ represents different types of system entities (\eg process, file, or network entities), and $op$ represents the system activity performed by the system call (\eg reading file or spawning a new process). 
Based on the types of the objects, system audit events are categorized as \emph{process events}, \emph{file events}, and \emph{network events}. 
\emph{Process events} record the operations of processes, such as execve. 
\emph{File events} record the operations on files, such as files read, write, and rename. 
\emph{Network events} record the operations of network accesses, such as sending and receiving messages from sockets. (See Table~\ref{tab:events} for more details.)

\subsection{Provenance Analysis (\pa)}
\label{subsec:causality-analysis}

\pa~\cite{backtracking,backtracking2,taser,intrusionrecovery,liu2018priotracker} analyzes the auditing events to infer their dependencies and present the dependencies as a directed graph, called a provenance graph (\pg).
In a \pg, a node is a system entity, such as a process, a file, or a network connection.
An edge represents a system audit event, and its direction indicates the direction of data flow (from $sub$ to $obj$ or vice-versa).
An edge is associated with event properties that are critical for security analysis (\eg data amount) and a time window that indicates the start time and the end time of the event.
Given a POI event, \pa starts from the POI event, and searches for other qualified events to form the \pg.
Formally, in the \pg $G(E,V)$, a node $v \in V$ represents a system entity, \ie a process, a file, or a network connection.
An edge $e(u, v) \in E$ indicates a system call event involving two entities $u$ and $v$ (\eg file read), and its direction (from $u$ to $v$) indicates the direction of data flow.
Each edge is associated with a time window, and $ts(e)$ and $te(e)$ are used to represent the start time and the end time of $e$, respectively.
Formally, in the \pg, for two events $e_1(u_1, v_1) $ and $e_2(u_2, v_2)$, there exists dependency between $e_1$ and $e_2$ if $v_1 = u_2$ and $ts(e_1) < te(e_2)$.
\pa performs two types of search on the system audit events: 
(1) \textit{backward \pa} that searches backward in time to find all the events that have causal dependencies on the POI event,
and (2) \emph{forward \pa} that searches forward in time to find all the events that the POI event has causal dependencies with.


\section{Overview}
\label{sec:overview}


\begin{figure*}[t]
    \centering
    \includegraphics[width=\textwidth,clip]{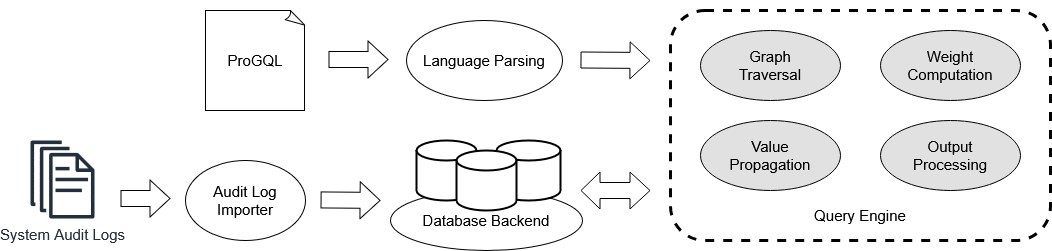}
    \caption{Overview of \tool framework}
    \label{fig:overview}
\end{figure*}

Our \tool framework consists of 3 major modules, as shown in \cref{fig:overview}.
The data importer module takes system auditing logs as input, and performs batch insertion into the database backend. 
The language parsing module takes a \tool query as input, parses the query text, and extracts the query context that contains all the required information for query execution.
The query engine is the core module of the \tool system, which executes the \tool query based on the extracted query context and searches the system auditing logs stored in the database backend to generate the desired \pg. 
The query engine consists of four components:
\textcircled{1} The graph traversal component finds the POI record in the database backend and performs graph search based on the query context.
In particular, incremental graph search is adopted to optimize the memory footprint and minimize search scope.
\textcircled{2} The weight computation component applies the weight computation function defined in the query context for each edge.
\textcircled{3} The value propagation component propagates the values defined in the query context for each node.
\textcircled{4} The graph merge component performs union or intersection of the graphs defined in the query context and outputs the processed graph as the \pg.

\section{Design of \tool}
\label{sec:approach}


In this section, we present the design details of each module shown in Figure~\ref{fig:overview}.

\begin{table}[!t]
	\centering
	\caption{Representative attributes of system entities}\label{tab:entity-attributes}

		\begin{tabular}{ll}
			\hline
			\multicolumn{1}{c}{\textbf{Entity}}		&\multicolumn{1}{c}{\textbf{Attributes}}\\\hline
			File				&id, path/name\\\hline
			Process			&id, name, pid\\\hline
			Network 	& id, src_ip, src_port, dst_ip, dst_port \\\hline
		\end{tabular}


	\vspace*{1ex}
\end{table}
 \begin{table}[!t]
	\centering
	\ra{1.2}
	\caption{Representative attributes of system events}
	\label{tab:event-attributes}
		\begin{tabular}{l|l}
			\hline
			\textbf{Operation}		& read, write, rename, create_object,\\
   &execute, clone, recvmsg, sendmsg  \\
			\textbf{Time}		& starttime, endtime\\
			\textbf{Misc.}		& ID, src_entity_ID, dst_entity_ID, amount\\\hline
		\end{tabular}

\end{table}

\eat{
\begin{table}[!t]
	\centering
	\caption{Representative attributes of system events}\label{tab1:event-attributes}
	\begin{adjustbox}{width=0.48\textwidth}
		\begin{tabular}{|l|l|}
			\hline
			Operation		& read/write,readv/writev,rename,modify_file_attributes,create_object,Execute/Execve,Clone,recvmsg/sendmsg,recvfrom/sendto,read_socket_params/write_socket_params,connect,\\\hline
			Time/Sequence		& start time/end time, \\\hline
			Misc.		& id, src_entity_id, dst_entity_id, amount\\\hline
		\end{tabular}
	\end{adjustbox}

		\vspace*{1ex}
\end{table}
}

\subsection{Audit Log Importer}
This module accepts system audit logs as input, constructs the data model for the events recorded in the logs, and performs batch insertion to import the modeled system events into multiple databases.

\noindent\textbf{Data Model}.
Existing works~\cite{backtracking,backtracking2,liu2018priotracker,gao2018aiql,gao2018saql,mcitracking} have indicated that on most modern operating systems (Windows, Linux, and OS X), system entities in most cases are files, processes, and network connections. Thus, in our data model, we consider system entities as files, processes, and network connections. We consider a system event as the interaction between two system entities represented by the tuple $\langle$subject, operation, object$\rangle$ where subject
represents a process entity, objects represent different types of
system entities, and
operation represents the system activity performed by the system
call (e.g., reading a file or spawning a new process).  We categorize system events into three types based on their object entities, namely file events, process events, and network events.
Both entities and events have critical security-related attributes shown in Tables~\ref{tab:entity-attributes} and~\ref{tab:event-attributes}. The attributes of entities include the properties to support various security analyses (\eg file path/name, process name, and ip addresses). The attributes of events include the source entity and the target entity, event origins (i.e., event id, starttime, endtime), and operations (\eg read/write).

\noindent\textbf{Batch Insertion}.
As system audit logs contain a huge amount of events~\cite{loggc,reduction,reduction2,reduction3,reduction4}, performing a \pa often requires scanning hundreds of millions of events. 
Thus, our importer adopts batch insertion to improve the performance of data insertion.
By doing so, we avoid lots of individual network message round-trips and other per-statement inefficiencies. 
To support near-real-time processing, we adopt a tiered-storage solution in our experiments. For hot data (most recent 1-3 days), we use SSD to support efficient batch insertion and gradually move the cold data to HDD. 
As system audit logs exhibit strong temporal/spatial properties~\cite{gao2018aiql,gao2018saql}, the data can be easily partitioned across different days and hosts. 
Such partitions automatically separate hot data and cold data and enable efficient data moving.

\subsection{\tool Language}


\rev {\tool introduces novel language constructs, which can express (1) dependency-based graph search that iteratively expands from already discovered edges by evaluating edge-dependent constraints to identify qualified adjacent edges, (2) assign weights to edges based on dynamically found edges and their adjacent edges, (3) propagate scores through weighted edges, and (4) merge multiple \pgs directly within a query using union and intersection, enabling analysts to combine backward and forward analyses or cross-host results without external post-processing. These are required features that are not supported by existing languages but necessary for \pa on system audit logs.

\label{appendix:grammar}

\autoref{bnf:parser} shows the representative BNF grammar of \tool langauge. 

\begin{BNF}
\footnotesize
\begin{mdframed}

\setlength{\grammarparsep}{-2pt} 
\setlength{\grammarindent}{10em} 

\begin{grammar}

<ProGQL> ::= sp? <ProGQLQuery> (sp (`union' | `intersect') sp `(' <ProGQLQuery> `)')* (sp? `;')? sp? EOF

<ProGQLQuery> ::= sp? <SingleQuery> (sp (`intersect' | `union') sp <WithQuery>)?

<SingleQuery> ::= ((sp? <Match>) (sp (<Bfs> | <Dfs>))? sp <Yield> sp <Return>)
    | ( ((sp? <Match>) (sp (<Bfs> | <Dfs>) sp <Yield> sp <Unwind>)? sp <Set> (sp <WithQuery>)?)+ (sp <Yield>)? SP <Return> )
               
<Match> ::= `match' sp? <Pattern> (sp? <Where>)?

<Pattern> ::= (<PatternPart> (sp? `,' sp? <PatternPart>)*) | <IdInColl> | <Expr>

<Expr> ::= `NOT' sp? <CompExpr> | (sp? <CompExpr> (`AND' | `OR') sp? <CompExpr>);

<CompExpr> ::= sp? <ArithmeticExpr> (sp? PartialCompExpr)*

<PartialCompExpr> ::= (`=' <ArithmeticExpr> | sp? <TraversalExpr>) | (`\textless \textgreater' <ArithmeticExpr> | sp? <TraversalExpr>) | (`\textless' <ArithmeticExpr> | sp? <TraversalExpr>) | (`\textgreater' <ArithmeticExpr> | sp? <TraversalExpr>) | (`\textless=' <ArithmeticExpr> | sp? <TraversalExpr>) | (`\textgreater=' <ArithmeticExpr> | sp? <TraversalExpr>)

<TraversalExpr> ::= sp? (<Max> | <Min>)

<ArithmeticExpr> ::= sp? <PropExpr> (((`*' | `/' | `+' | `-') <PropExpr>)+)?;

<Max> ::= `max' sp? `(' sp? <Collect> sp? `)'

<Min> ::= `min' sp? `(' sp? <Collect> sp? `)'

<Collect> ::= `collect' sp? `(' sp? <IdInColl> sp? `|' sp <PropExpr> sp? `)'

<PropExpr> ::= sp? <Atom> (sp? <PropLookup>)*

<PropLookup> ::= `.' sp? <PropKeyName>


<IdInColl> ::= sp? <Var> sp `in' <Expr>

<Bfs> ::= `bfs' sp? `(' sp? <Var> sp `in' (sp? <Backward>|sp? <Forward>) sp? `|' sp <Match> sp? `)'

<Backward> ::= `backward' sp? `(' <Var> `)'

<Forward> ::= `forward' sp? `(' <Var> `)'

<Dfs> ::= `dfs' sp? `(' sp? <Var> sp `in' (sp? <Backward>|sp? <Forward>) sp? `|' sp <Match> sp? `)'

<Where> ::= `where' sp <Expr> (sp <Order>)? (sp <Limit>)?

<Order> ::= `order' sp `by' sp <SortItem> (`,' sp? <SortItem>)*

<SortItem> ::= sp? <PropExpr> (sp? (`asc' | `desc'))?

<Limit> ::= `limit' sp <NumberLiteral>

<Uwind> ::= `uwind' sp <Var> sp `as' sp <Var>

<Yield> ::= `yield' sp <Var>

<Atom> ::= sp? (<Literal> | <Var> | <FuncInvoc> | <ParenExpr>)

<ParenExpr> ::= `(' sp? <Expr> sp? `)'

<Set> ::= `set' sp? <SetItem> (`,' <SetItem>)*

<SetItem> ::= (sp? <PropExpr> sp? `=' (sp? <Expr> | sp? <Projection>))|(sp? <PropExpr> sp? `=' (sp? <Expr> | sp? <Reduce>))

<Projection> ::= `projection' sp? `(' (sp? <Expr> (`,')?)+ sp? `)'

<Reduce> ::= `reduce' sp? `(' <Var> sp? `=' (sp? <NumberLiteral>) sp? `,' sp? <IdInColl> sp? `|' (sp? <Expr>) sp? `)'

<Return> ::= `return' sp <Var>

<WithQuery> ::= `with' sp <Var> ((sp? `=' sp? `(' sp? <Match> sp? `)' sp (<Bfs>|<Dfs>) sp <Yield> sp <Return>) | (sp <Where>))

<FuncInvoc> ::= <FuncName> sp? `(' <Expr> `)'

<FuncName> ::= `count' | `dst' | `src' | `out' | `in' | `abs' | `ln' |...
\end{grammar}

\end{mdframed}

\caption{Representative BNF grammar of \tool}
\label{bnf:parser}
\end{BNF}
 }
The grammar specifies both the adapted constructs (e.g., MATCH, WHERE, RETURN) and the new operators introduced in \tool, which include:

\noindent\textbf{Constrained Graph Search}.
\pa constructs tailored graphs based on the constraints (\eg recursively filtering edges based on timestamps). 
To express a constrained graph search, \tool defines rules \emph{$\langle$Bfs$\rangle$} and \emph{$\langle$Dfs$\rangle$} \rev{combined with \emph{$\langle$Backward$\rangle$} and \emph{$\langle$Forward$\rangle$} to define traversal strategy and direction. 
For example, }
\incode{backward(f)} traces back from the node \incode{f}, and \incode{forward(entry)} traces forward from the node \incode{entry}. 
The search can be constrained using the rules \emph{$\langle$Match$\rangle$} and \emph{$\langle$Where$\rangle$}\rev{, which filter candidate edges or nodes at each expansion step}.

The rule \emph{$\langle$Match$\rangle$} uses \emph{$\langle$Pattern$\rangle$}, which can be a combination of multiple \emph{$\langle$PatternPart$\rangle$} separated by commas to specify an edge pattern for finding qualified edges (Line 1). 
Or it can be a single identifier within a collection (\emph{$\langle$IdInColl$\rangle$}) (Line 8: \incode{n in nodes(r)}), or an expression \emph{$\langle$Expr$\rangle$} (Line 2: \incode{v=dst(r)}) to specify qualified nodes.  
The rule \emph{$\langle$Where$\rangle$} specifies the search conditions using the rule \emph{$\langle$Expr$\rangle$}, which is defined to express logical and arithmetic operations, including negation, conjunction (AND), disjunction (OR), comparison operations, traversal expressions (\emph{$\langle$TraversalExpr$\rangle$}), and arithmetic expressions (\emph{$\langle$ArithmeticExpr$\rangle$}) on edge properties (\emph{$\langle$PropExpr$\rangle$}).

The rule \emph{$\langle$TraversalExpr$\rangle$} \rev{further enhances expressiveness by allowing aggregation over traversals. Specifically, \emph{$\langle$TraversalExpr$\rangle$}} uses the rule \emph{$\langle$Max$\rangle$} or the rule \emph{$\langle$Min$\rangle$} to compute the maximum or minimum value of the selected property, \rev{and \emph{$\langle$Collect$\rangle$} to gather sets of properties, enabling iterative constrained search.
For example, the condition (Line 2: \incode{r.starttime < max(collect(vout IN out(v) | vout.endtime)))} is used to retain only the edges whose start time are earlier than the max end time of all the outgoing edges.}
The rules \emph{$\langle$Atom$\rangle$}, \emph{$\langle$PropLookup$\rangle$}, \emph{$\langle$FuncInvoc$\rangle$}, \emph{$\langle$FuncName$\rangle$} and \emph{$\langle$ParenExpr$\rangle$} \rev{allow flexible combinations of edge properties and functions.}

\noindent\textbf{Weight Computation}.
The graph generated by the constraint graph search contains the contextual information of an attack.
However, this graph can be gigantic, typically containing $>$ 100,000 edges. 
As a result, it is difficult for security analysts to find the edges that are critical to the attack. To address this issue, \tool extends the rules \emph{$\langle$Unwind$\rangle$}, \emph{$\langle$Set$\rangle$} and defines a new rule \emph{$\langle$Projection$\rangle$} to compute edges weights so that edge weights based on critical security properties can be defined to filter irrelevant edges~\cite{hassan2019nodoze,depimpact}.
The rules \emph{$\langle$Unwind$\rangle$} and \emph{$\langle$Set$\rangle$} are used to specify edge weight computation based on edge properties such as data amount, in-degrees, out-degrees, and time, which are specified in the rule \emph{$\langle$Projection$\rangle$} (Lines 4).
Recent studies~\cite{hassan2019nodoze,liu2018priotracker,depimpact} show that combinations of multiple edge properties work well for a wide range of attacks.
In particular, the rule \emph{$\langle$Projection$\rangle$} indicates to use a projection function that projects the selected property values into a single-dimensional weight such as LDA~\cite{Mika99fisherdiscriminant} to obtain the edge weight score. 
An edge with a higher dependency weight score implies more relevance to the POI event, and is more likely to be a critical edge.

\noindent\textbf{Value Propagation}.
To reveal attack entries, \tool extends the rules \emph{$\langle$Match$\rangle$}, \emph{$\langle$Set$\rangle$} and \emph{$\langle$Reduce$\rangle$} to compute nodes' values based on edge weights, which is used to model the node's impact on the POI event. In Query ~\ref{query:pwd}, the score propagation scheme computes the impact score of a node as a weighted sum of its children's impact scores.

\noindent\textbf{Graph Merge}.
Existing studies~\cite{depimpact,backtracking2} show that multiple \pa can be used to connect attack behaviors across hosts, or perform more effective filtering on irrelevant edges.
\rev{\tool defines graph merge composition through the rules \emph{$\langle$ProGQL$\rangle$} and \emph{$\langle$ProGQLQuery$\rangle$}. The rule \emph{$\langle$ProGQL$\rangle$} specifies that a ProGQL program may consist of one or more queries, optionally combined with the keywords \incode{union} or \incode{intersect}. Each merge operation takes as input a \emph{$\langle$ProGQLQuery$\rangle$} enclosed in parentheses, allowing the results of multiple queries to be unified into a single output graph. The rule \emph{$\langle$ProGQLQuery$\rangle$} further refines this by allowing each query to contain one or more \emph{$\langle$SingleQuery$\rangle$} statements, optionally followed by additional merge clauses (\incode{union} or \incode{intersect}) together with a \emph{$\langle$WithQuery$\rangle$}. This enables analysts to express multi-stage query pipelines where intermediate graphs are named and then combined. \emph{$\langle$WithQuery$\rangle$} provides a hook to bind intermediate results for further processing. Its full semantics are illustrated in the Query~\ref{query:pwd} (e.g., ranking and selecting entry nodes before forward search).}

\noindent\textbf{Query Example}.
Query~\ref{query:pwd} shows a \tool query for investigating password crack. After successfully penetrating host1, an attacker downloads and executs malicious scripts to identify additional victim hosts, such as host2. The attacker then cracks host2's password data, copys it back to host1, and compresses it for further malicious activity.

\vspace{1em} 
\begin{lstlisting}[caption={\tool Query for Password Crack},captionpos=b,label={query:pwd}]
MATCH (p:Process)-[st:FileEvent{optype:"write"}]->(f:File{name:"/tmp/passwords.tar.bz2", hostid:"1"})
        BFS (r IN backward(f) | MATCH v=dst(r) WHERE r.starttime<max(collect(vout IN out(v) | vout.endtime))) YIELD g1
        UNWIND g1 AS e
        SET e.weight=projection(1/(abs(r.amount-st.amount)+0.0001),ln(1+1/abs(r.endtime-st.endtime)),count(out(v))/count(in(v)))
        MATCH u=src(e) SET u.rel=reduce(sum = 0, o IN out(u) | sum+o.weight*dst(o).rel)
        RETURN g1
intersect
WITH entry = (MATCH n in nodes(r) WHERE count(in(n))=0 ORDER BY n.rel DESC LIMIT 15)
        BFS (re IN forward(entry) | MATCH u=src(re) WHERE re.endtime>min(collect(uin IN in(u) | uin.starttime)) and re.starttime<1724731846719889370) yield g2
        RETURN g2

UNION
(MATCH (p:Process)-[st:NetworkEvent{id:100005}]->(f:Network{srcip:"192.168.1.128/32",dstip:"192.168.1.131/32",hostid:"2"})
        BFS (r IN backward(f) | MATCH v=dst(r) WHERE r.starttime<max(collect(vout IN out(v) | vout.endtime))) YIELD g3
        UNWIND g3 AS e
        SET e.weight=projection(1/(abs(r.amount-st.amount)+0.0001),ln(1+1/abs(r.endtime-st.endtime)),count(out(v))/count(in(v)))
        MATCH u=src(e) SET u.rel=reduce(sum = 0, o IN out(u) | sum+o.weight*dst(o).rel)
        RETURN g3
intersect
WITH entry = (MATCH n in nodes(r) WHERE count(in(n))=0 ORDER BY n.rel DESC LIMIT 15)
        BFS (re IN forward(entry) | MATCH u=src(re) WHERE re.endtime>min(collect(uin IN in(u) | uin.starttime)) and re.starttime<1724731846712161377) yield g4
        RETURN g4)

\end{lstlisting}

In Query~\ref{query:pwd}, for instance, the union of two hosts' output \pgs is computed (Line 12) to help reassemble an attack scenario that involves password cracking and data exfiltration across hosts.
For each host, a backward graph search and a forward graph search are employed and their output graphs are intersected to form the host's \pg.
Specifically, \tool uses the rule \emph{$\langle$WithQuery$\rangle$} \rev{to bind intermediate subgraphs and identify entry nodes for further exploration. In Query~\ref{query:pwd}, \incode{WITH entry = (...)} names a subgraph expression that can be reused in later traversals or merge operations. Entry nodes are selected with \incode{nodes(r)} and filtered using \incode{count(in(n))=0}, which captures provenance boundaries where external influence first enters the system. Candidate entry nodes are then ranked by their propagated impact scores using \emph{$\langle$Order$\rangle$}, and \emph{$\langle$Limit$\rangle$} is applied to restrict forward exploration to the top-k nodes. Once these entry nodes are chosen, a forward \emph{$\langle$Bfs$\rangle$} is performed from them to construct a new provenance subgraph. This forward \pg is then combined with the backward \pg using \incode{intersect}, which captures the overlap between the two searches and yields a focused explanation of the attack chain.}



\subsection{Query Execution Engine}
\label{subsec:engine}
The query engine executes the query context generated by the language parser and outputs a \pg. It consists of four components: \textcircled{1} graph traversal, \textcircled{2} weight computation, \textcircled{3} value propagation, \textcircled{4} \rev{graph merge}.

In the graph traversal component, the engine synthesizes database queries to perform graph searches with the help of the database backend. 
In the weight computation component, 
\rev{when a query specifies edge re-materialization and weighting (e.g., \incode{UNWIND g1 AS e} followed by \incode{SET e.weight=projection(...)})},
the engine first applies an edge merge technique to reduce the size of the \pg that is generated by the graph search. Then it will compute the edge's weight score based on the edge features defined in the projection function and employ a discriminate feature projection scheme based on LDA to project the selected features into a single-dimensional weight, so that critical edges can be better revealed. 
In the value propagation component, \rev{when a query specifies score assignment and propagation (e.g., \incode{MATCH u = src(e)} followed by \incode{SET u.rel = reduce(...)})}, the engine applies a weighted score propagation scheme to propagate the impact score from the POI node to all other nodes in the \pg. 
In the output processing component, if \incode{union} or \incode{intersect} is declared in the query, the engine will merge multiple generated \pgs generated and output a final \pg.
We next describe each phase in detail.

\noindent\textbf{Component \textcircled{1}: Graph Traversal}.
Based on the \tool query semantics, the engine synthesizes database queries (SQL for the relational database, nGQL for NebulaGraph, and traversal API calls for Neo4j), which first identify the starting node of the traversal (POI node that is extracted from the POI event defined in the graph search query, or attack entry node defined in the \incode{WITH} query) and add it to a queue. 
Then \tool performs incremental backward/forward search from the starting node, without the need to load all events that happen before the starting node.
Such incremental search greatly reduces the number of events to be checked as the number of events to be loaded is usually huge (easily exceeding 10 million events) and many events do not have dependency paths that lead to the starting node.
Note that this incremental search will not be possible if a database backend is not adopted and the events are not indexed.

With the starting node in the queue, the engine pop ups the node from the queue and constructs database queries based on the search constraints (\eg the \incode{MATCH} query in Line 2 in Query~\ref{query:pwd}) to find eligible incoming/outgoing edges. If any edges are found, their source/sink nodes are added to the queue for further search, and the engine continues to pop up the node from the queue.
This process repeats until the queue is empty or the search exceeds the resource limits.
\cite{progqlbackwardbfs} shows the detailed implementation of a backward BFS search from a POI node.  

\noindent\textbf{Component \textcircled{2}: Weight Computation}.
Weight computation consists of three major steps:
\begin{itemize}[noitemsep, topsep=1pt, partopsep=1pt, listparindent=\parindent, leftmargin=*]
\item \textit{Edge Merge}: the engine reduces the \pg size by merging parallel edges between two nodes in the \pg. \pa often results in a \pg with numerous parallel edges connecting two nodes, which arises from the operating system's tendency to distribute data across several system calls upon completing read/write tasks, such as file operations~\cite{backtracking,loggc}. 
The engine addresses this issue by merging edges between nodes when the time differences are below a specified threshold, following existing works~\cite{reduction,depimpact}. 
\item \textit{Weight Evaluation}: the engine computes the weights of the edges using the features defined in the \incode{projection} function.
The engine extracts features for each edge from the \incode{projection} function, along with the corresponding arithmetic expressions for weight calculation. Subsequently, an arithmetic evaluator is employed to calculate the weight of each individual feature.
For example, in Query ~\ref{query:pwd},  \incode{ln(1+1/abs(r.endtime-st.endtime))} represents a temporal weight that models the temporal relevance of an edge $r$ to the POI event $st$. 
Similarly, \incode{1/(abs(r.amount-st.amount)+0.0001} represents a data flow weight that models the data flow relevance of $r$ to $st$.

\item \textit{Weight Normalization}: the engine normalizes the weights of all outgoing edges for each node, and thus the final weights are local weights for each node. 
It is an optional step, and it can potentially mitigate the weight degradation for the edges that are far from the POI event.
To do so, the engine applies a clustering algorithm such as KMeans~\cite{Arthur:2007:KAC:1283383.1283494} or DBScan~\cite{schubert2017dbscan} to separate edges into critical edges group and non-critical edges group. 
After the clustering, the engine employs a discriminative feature projection scheme (\eg LDA) to compute a final weight based on the individual features' weights. Finally, the engine normalizes an edge $e(u,v)$'s final weight by the sum of weights of all outgoing edges of the source node $u$.


\end{itemize}

\noindent\textbf{Component \textcircled{3}: Value Propagation}.
The engine propagates values from the POI node to all other nodes backward along the weighted edges following the \incode{reduce} function definition in the query. For example, according to \incode{reduce(sum = 0, o IN out(u) | sum+o.weight*dst(o).rel)}, an arithmetic evaluator is used to calculate a node's impact score by taking the weighted sum of impact scores of its child nodes. 
Note that the value propagation is recursive. In each iteration, the node's impact score is updated based on the impact scores of its child nodes. 
The iterative process continues until the impact scores converge to stable values. The engine will terminate the propagation when the aggregate difference between the current iteration and the previous iteration is smaller than a threshold (\eg $1e\mbox{-}13$), as it indicates that the propagated values become stable after iterations.


\noindent\textbf{Component \textcircled{4}: \rev{Graph Merge}}.
This component merges the \pgs \rev{produced by each sub-query and supports query-level merge constructs such as \incode{union(g1, g2)}, which combines the nodes and edges of two subgraphs, and \incode{intersect(g1, g2)}, which retains only the overlapping portions. To perform merging efficiently, the engine constructs an edge signature by leveraging the unique identifiers of both the source and sink nodes; this signature enables fast decisions on whether an edge should be included, greatly improving performance when combining multiple \pgs. Finally, the merged \pg is transformed into the output format specified in the \tool query.}

\section{Evaluation}
\label{sec:eval}
We built \tool ($\sim$32K lines of code for the language and $\sim$10K lines of code for log parser) on top of both relational database PostgresSQL 9.5~\cite{postgresql}, MyRocks (\aka Facebook MySQL 5.6)~\cite{myrocks}, Mariadb 10.4.19~\cite{mariadb} and graph database Neo4j 3.5.11~\cite{neo4j}, Nebula 3.3.0~\cite{nebula} and evaluate \tool using both the attack cases constructed based on the known exploits~\cite{exploitdb,liu2018priotracker,kwon2018mci,reduction} and the attack cases collected by the DARPA Transparent Computing (TC) program~\cite{tc}. 
We constructed 14 \tool queries to investigate these attacks, demonstrating the effectiveness of \tool in searching attack behaviors for diverse attacks. 
In the evaluations, we aim to answer the following research questions:

\begin{itemize}[noitemsep, topsep=1pt, partopsep=1pt, listparindent=\parindent, leftmargin=*]
\item \textbf{RQ1}: How effective is \tool in expressing different \pa to generate \pgs for finding attack behaviors? 
\item \textbf{RQ2}: How does \tool improve the efficiency and accuracy of identifying attack patterns in audit logs compared to the most popular graph query language, Cypher?
\item \textbf{RQ3}: How effective is \tool in revealing attack steps for advanced cyberattacks? How does \tool compare with other state-of-art (SOTA) techniques in attack investigation? 
\item \textbf{RQ4}: \rev{How flexible can \tool support different weight computations?}

\item \textbf{RQ5}: How do different backend databases impact \tool's runtime performance? 

\end{itemize}

\subsection{Evaluation Setup}
\label{subsec:evalsetup}
We deployed \tool on a server with an Intel(R) Xeon(R) CPU E5-2637 v4 (3.50GHz), 256GB RAM running 64bit Ubuntu 16.04.7. Neo4j and Nebula databases are configured by importing system entities as nodes and system events as relationships. 
To evaluate the effectiveness and performance of \tool, we deployed Sysdig~\cite{sysdig} on Linux hosts to collect system auditing events and performed a broad set of attack behaviors including 6 attacks based on commonly used exploits and 3 multi-host intrusive attacks based on the Cyber Kill Chain framework~\cite{cyberkillchain} and CVE reports~\cite{cve}.
The DARPA datasets include system audit logs collected from 4 hosts with different OS systems.

\myparatight{Batch Insertion}
Batch insertion can support 3000 events/second (supporting 50-100 monitored hosts) using HDD, and improve to 8000 events/second using SSD. 
Modern DBs support incremental index building and query speed is improved significantly by 2-3 times using SSD. 
Thus, \tool can support near-real-time queries.

\myparatight{Database Backend Setup}
We developed a tool to parse Sysdig logs and DARPA released logs and loaded them into five databases: PostgresSQL, MyRocks, Mariadb, Neo4j, and Nebula.
We then deploy \tool to use these five databases as five types of database backends.
For evaluation, we used datasets containing a total of 140,523 entities and 28,088,979 events, in addition to the separate DARPA datasets with 46,756,662 and 78,219,245 events, respectively.
We next describe these attacks in detail. 

\subsubsection{Attacks Based on Commonly Used Exploits}
\label{subsub:benign-cases}
These 6 attacks are used in prior work's evaluations~\cite{exploitdb,liu2018priotracker,kwon2018mci,reduction},
which consist of the following scenarios: 
\begin{itemize}[noitemsep, topsep=1pt, partopsep=1pt, listparindent=\parindent, leftmargin=*]
    \item \textit{Wget Executable}: A vulnerable server allows the attacker to download executable files using wget. 
    The attacker downloads Python scripts and executes the scripts to write some garbage data to a user's home directory.
    \item \textit{Illegal Storage}: A server administrator uses wget to download suspicious files to a user's directory.
    \item \textit{Hide File}: The goal of the attacker is to hide malicious files among a user's normal files. 
    The attacker downloads the malicious file and hides it by changing its file name and moving it to a user's home directory.
    \item \textit{Steal Information}: The attacker steals the user's sensitive information and writes the information to a hidden file.
    \item \textit{Backdoor Download}: A malicious insider uses the ping command to connect to the malicious server, and then downloads the backdoor script from the server and renames the script to hide it.
    \item \textit{Annoying Server User}: 
    The annoying user logs into other users' directories on a vulnerable server, and runs the backdoor script downloaded from the malicious server using ping command to write some garbage data to other users' directories. 
\end{itemize}

\subsubsection{Multi-host Intrusive Attacks}
\label{subsubsec:attack-cases}

This multi-host intrusive attack encapsulates key characteristics found in the Cyber Kill Chain framework~\cite{cyberkillchain} and Common Vulnerabilities and Exposures (CVE)~\cite{cve}. In these three attack scenarios, the attacker leverages an external host, designated as the C2 (Command and Control) server, to execute penetration, disseminate malware, steal data, and establish the persistent connection. The target host compromised by the attack is referred to as the Victim host.

\myparatight{Attack1: Password Crack After Shellshock Penetration}
The Shellshock vulnerability  was a critical flaw discovered in the GNU Bash shell. It allowed attackers to execute arbitrary commands on a targeted system through specially crafted environment variables. When Bash processes these variables, the attacker’s code gets executed. This attack exploited the Shellshock vulnerability to infiltrate a vulnerable host (host1). \rev{Upon successful compromise, the attacker establishes a reverse shell connection to remotely control host1. In this stage, the attacker generally takes a series of stealthy reconnaissance maneuvers. Among those, we emulate the password cracking attack. The attacker downloads and executes a malicious script \incode{gather_password.sh}. This script identifies victim hosts (i.e., host2) and downloads another malicious script \incode{crack_passwd.sh}, transfers it to host2 and executes it. \incode{crack_passwd.sh} then downloads a series of files, including a malicious payload \incode{libfoo.so} from the attack server. \incode{libfoo.so} cracks passwords on the victim host. The resulting \incode{password_crack.txt} file contains plaintext passwords, which is then transferred to host1 and compressed as the \incode{/tmp/passwords.tar.bz2}. This file serves as a consolidated package of sensitive information, ready for exfiltration.}

\myparatight{Attack 2: Data Leakage}
\rev{ After Shellshock Penetration, the attacker attempts to steal all the valuable assets from the host. This stage mainly involves the behaviors of local and remote file system scanning activities, copying and compressing of important files. The attacker initiated the second stage of the attack by downloading the script \incode{leak_data.sh} from their server. The script was then transferred to a compromised host (host2) using scp and executed. The \incode{leak_data.sh} script bundled specific files, including hidden files and sensitive system files, into a tarball. This tarball was then compressed as \incode{leaked.tar.bz2} and exfiltrated to the attacker’s designated server for further use.
}

\myparatight{Attack 3: VPN Filter}
At this stage, the attacker targeted to establish a persistent connection between the victim hosts and the C2 server. The attacker utilized VPNFilter malware, which infected millions of IoT devices by exploiting a number of known or zero-day vulnerabilities. The attacker first downloaded \incode{vpn_filter.sh} script to host1, the script then was transferred to host2 and executed. \incode{vpn_filter.sh} changed to the /tmp directory on host2, downloaded \incode{vpnfilter} from the attacker's server, made it executable, and then ran it with C2 server IP address. This established a persistent connection between host2 and the C2 server.

\subsubsection{DARPA TC Attack Cases}
The datasets provided by the DARPA TC program feature instances of attacks conducted on various operating systems. Our selection process, guided by DARPA's ground truth document, involved excluding failed attack cases and those targeting the Android system. The latter was omitted due to the constrained behaviors of mobile applications within the Android sandbox, rendering them unsuitable for our analysis. Additionally, we opted to exclude phishing email attacks, given their reliance on browser interactions, which result in limited traces within system audit logs.

Ultimately, our focus narrowed to five distinct attacks that target diverse operating systems such as Linux and Windows, exploiting vulnerabilities such as the Firefox backdoor and browser extensions. Notably, these attack cases unfold over extended periods, exemplified by Theia data encompassing logs spanning 8 days.

\subsubsection{Obtaining Ground Truth for the Attacks}
Within our best efforts, we manually ensured that the critical events that represent attack steps were identified based on the knowledge of the attacks performed by us and DARPA attack descriptions in these \pgs.
\cref{tab:attacks} shows the statistics of all the 14 attacks. 
Columns ``Critical Event'' shows the critical edges that represent attack steps of the \pgs. 
``System Entity'' and ``System Event'' show the number of entities and events that were imported into the database by our Data Importer. 
On average there are 3,562,996 entities and 18,556,587 events (with the max being 44,455,147 entities and 64,422,779 events). 
Loading them into the memory and performing the \pa are super expensive, which motivates the database approach provided by \tool.

\begin{table}[!t]
	\centering
	\caption{Statistics of all the 14 attacks}
        \label{tab:attacks}
	\resizebox{\linewidth}{!}{%
        \begin{tabular}{lrrr}
        \toprule
        \multicolumn{1}{c}{\textbf{Attack}}     & \multicolumn{1}{c}{\textbf{Critical Event}} & \multicolumn{1}{c}{\textbf{System Entity}}  & \multicolumn{1}{c}{\textbf{System Event}}\\ \midrule
        Wget Executable              & 12           & 134,066   & 27,836,522  \\     
        Illegal Storage            & 6    & 134,066   & 27,836,522          \\
        Hide File  & 8  & 134,066   & 27,836,522   \\
        Steal Information  & 7  & 134,066   & 27,836,522   \\
        Backdoor Download  & 6  & 134,066   & 27,836,522   \\
        Annoying Server User  & 12  & 134,066   & 27,836,522   \\
        Password Crack  & 37  & 6,457   & 252,457   \\
        Data Leakage  &17 &6,457 & 252,457  \\
        Vpn Filter  & 16 &6,457 &252,457 \\
        Theia Case 1  & 7  & 1,487,424   & 8,704,250    \\
        Theia Case 3  & 4  & 1,487,424   & 8,704,250    \\
        Fivedirections Case 1  & 4  & 814,091   & 5,092,216    \\
        Fivedirections Case 3  & 2  & 814,091   & 5,092,216    \\
        Trace Case 5  & 3  & 44,455,147   & 64,422,779 \\ \midrule 
        \textbf{Average} & 10 & 3,562,996 & 18,556,587\\ \bottomrule
        \end{tabular}
	}
\end{table}

\myparatight{Evaluation Metrics} 
To measure the effectiveness of \tool, we counted false positives (detected edges that are not critical edges) and false negatives (missing edges that are critical edges) and showed them in \cref{tab:comparesize} Columns ``FP'' and ``FN''.

\begin{table*}[!t]
	\centering
	\caption{\pa executed by \tool for 14 Attacks}
 \label{tab:rq1}
	\resizebox{\linewidth}{!}{%
        \fontsize{4}{4}\selectfont
        \begin{tabular}{lrrrrrrr}
        \hline
        \multicolumn{1}{c}{\textbf{Attack}}     & \multicolumn{1}{c}{\textbf{\# Sub-Q}} & \multicolumn{1}{c}{\textbf{\# E of Backward \pa}} &
        \multicolumn{1}{c}{\textbf{Top Nodes} } 
        &\multicolumn{1}{c}{\textbf{\# E of Forward \pa}} &
        \multicolumn{1}{c}{\textbf{\# E of \pg}} &\multicolumn{1}{c}{\textbf{Memory (GB)}} &\multicolumn{1}{c}{\textbf{Exe. Time (s)}} \\ \hline
        Wget Executable     & 1   & 46,798    &  5     & 209   & 65 & 1.61  & 99  \\     
        Illegal Storage       & 1   &  34,242   &  3   & 46,956  & 212 & 1.94  & 128       \\
        Hide File  & 1   &  43,444   &  2   &  20,726 & 46 & 1.63  & 90   \\
        Steal Information  & 1   &  52,292   &  6   & 26,773  &50  & 2.01   & 140  \\
        Backdoor Download  & 1   &  46,454   &  2   & 128  & 47 & 1.44  & 93  \\
        Annoying Server User  & 1   & 2,046,735    &  3   &  158 & 46 & 15.07  &  629   \\
        Password Crack  & 2   & 20,349    &  30   & 27,797  & 282 & 1.94  & 23  \\
        Data Leakage  & 2   & 24,177   & 10    & 23,254  & 281 & 1.45  &  16 \\
        Vpn Filter  & 3   & 50,935  &  18  &  30,631 & 637 & 1.91  & 23  \\
        Theia Case 1  & 1   & 27,397    &  2   & 1,398  & 339 & 3.00  &  242  \\
        Theia Case 3  &  1  &  8   &   3  & 1,039,978  & 6 &  8.30 & 320  \\
        Fivedirections Case 1  & 1   &   319  &  3   &  32 & 14 &  1.46 &  56 \\
        Fivedirections Case 3  &  1  &  17   &  1   & 604,836  & 17 &  5.32 & 210  \\
        Trace Case 5  & 1   &  827   &  3   & 1,028,353  & 474 & 24.76  & 1,069 \\ \hline 
        \textbf{Average} &  1  &  171,000   &   7  & 203,659  & 180 & 5.13  & 224 \\ \hline
        \end{tabular}
        
	}
\label{tab:rq1}
\end{table*}

\subsection{RQ1: Expressing \pa for Attacks}
We compose 14 \tool queries for the 14 attacks and execute them in our deployed \tool framework to generate \pgs for performing attack investigation.  
\cref{tab:rq1} presents statistical insights into the generated \pgs.
Based on the existing studies~\cite{depimpact,mcitracking,sleuth}, \tool queries are formulated to comprise of \rev{1$\sim$3} sub-queries. 
In each sub-query, The first nested sub-query initiates from a given POI event and conducts backward \pa with the specified qualification constraints, aiming to identify all the potentially attack-related edges. 
This nested sub-query also computes edge weights based on three features (data amount, relative time, and ratio of incoming and outgoing edges) as suggested by the existing studies, and propagates values to each identified nodes as each node's impact score.
Note that the \tool language can easily incorporate any new edge-related feature as long as the system auditing events can provide the required data. 
Subsequently, in the second nested sub-query, we introduce a condition to choose the top $N$ entry nodes as the starting nodes to perform the subsequent forward \pa.
Finally, the \tool query merges the output \pgs from the all sub-queries using \incode{INTERSECT} / \incode{UNION}. 
The merged \pg preserves the nodes and edges that are highly relevant to both the POI event and the attack entries. 
All queries are available in our project website~\cite{progql}.

\cref{tab:rq1} presents the statistical insights of the generated \pgs, where the Column ``\# Edges of Backward \pa'' shows the number of edges in the \pg generated by the first sub-query, Column ``\# Edges of Forward \pa'' shows the number of edges in the \pg generated by the second sub-query, and Column ``\# Edges of Output \pg'' shows the number of edges in the final merged \pg.
We can observe that the \pg generated by the backward sub-queries is very large (averagely 171,000), while the final merged \pg has about 180 edges (merely 0.11\%).
This shows that \textit{multiple \pa collaborates with each other with the help of weight computation and value ranking is very effective in reducing irrelevant edges, which is consistent with the previous studies~\cite{backtracking2,depimpact}}.

\myparatight{Execution Performance and Memory Consumption}
\cref{tab:attacks} shows that the number of events to search is about 19 million on average, while the output \pg that are useful for attack investigation has the average size of 180, which is like ``finding needle in haystack''.
With the edge indexing and the incremental graph search, when deploys \tool on Neo4j, it can conduct such search using averagely 5GB memory and finish the search within 224 seconds, while existing works such as \depimpact~\cite{depimpact} that load all the data into the memory can easily consume up to 100GB memory (see \cref{subsec:rq2}) or cause time out. 
These results demonstrate the optimization brought by our query engine design and the improved scalability.

\myparatight{Top $N$ Ranked Nodes for Forward \pa}
Column ``Top Ranked Nodes'' shows the number of selected entry nodes for performing the forward \pa. 
With the help of the \tool language, security analysts can effortlessly customize this number by adapting the query (\incode{WITH entry = (MATCH n in nodes(r) WHERE count(in(n))=0 order by n.rel desc limit "N")}) to align with their domain knowledge and the observed outputs. 
In our evaluations, we keep increasing the value of $N$ until all critical edges are found or a larger $N$ will include too much irrelevant edges into the final \pg.

\begin{table*}[!t]
	\centering
	\caption{\pgs produced by \tool and Cypher (Neo4j)}
 \label{tab:comparesizecypher}
	\resizebox{\linewidth}{!}
 {
        \begin{tabular}{l|rrr|rrr|c|c|c|c}
\hline
\multicolumn{1}{c|}{\multirow{2}{*}{\textbf{Attack}}} &  
\multicolumn{3}{c|}{\textbf{Cypher (Neo4j)}}  & 
\multicolumn{3}{c|}{\textbf{ProGQL}}  &  
\multicolumn{1}{c|}{\textbf{Cypher (Neo4j)}} & 
\multicolumn{1}{c|}{\textbf{ProGQL}} &  
\multicolumn{1}{c|}{\textbf{Cypher (Neo4j)}} &  
\multicolumn{1}{c}{\textbf{ProGQL}}  
\\ \cline{2-7}
& \multicolumn{1}{c}{\textbf{FP}} & 
\multicolumn{1}{c}{\textbf{FN}}   & 
\multicolumn{1}{c|}{\textbf{Edges}} & 
\multicolumn{1}{c}{\textbf{FP}} & 
\multicolumn{1}{c}{\textbf{FN}}   & 
\multicolumn{1}{c|}{\textbf{Edges}} 
& \multicolumn{1}{c|}{\textbf{Execution Time(s)}} & 
\multicolumn{1}{c|}{\textbf{Execution Time(s)}} &  
\multicolumn{1}{c|}{\textbf{Memory Consumption(GB)}} &  
\multicolumn{1}{c}{\textbf{Memory Consumption(GB)}}  
\\ \hline

        Password Crack  & 21,481  &  0  & 21,518 & 245 &0 & 282 & 612 &23 &  5.53 & 1.94\\
        Data Leakage  & 21,539  &  0  & 21,556 & 264 & 0 & 281  &185 & 16 & 1.26 & 1.45\\
        Vpn Filter  & 24,817  &  0  & 24,827 &627 &0 & 637  &638 & 23& 2.09 & 1.91 \\
        \hline
        \textbf{Average}  & 22,612  &  0  & 22,634 & 379 & 0 &  400& 478 & 21& 2.96 & 1.77\\
        \hline
        \end{tabular}
	}
\end{table*}

\myparatight{Customized Filtering Constraints}
Unlike existing \pa techniques, as a domain-specific language, the \tool language allows security analysts to effortlessly incorporate additional constraints into the query for filtering irrelevant edges based on their domain knowledge or the observed outputs. 
In our evaluations, we added customized filters to filter out irrelevant system libraries files that are read-only and are not tampered with. 
While these libraries may vary in different systems, our queries can easily adapt to different systems by adjusting the filtering constraints.
Furthermore, we added temporal constraints specifically in the analysis of DARPA TC attacks, given the substantial size of their datasets. 
For instance, the Theia dataset spanning logs over 8 days. 
Without imposing time constraints, conducting a search from a POI event would undoubtedly result in an overwhelmingly large \pg. 
To manage this, we apply time constraints (\incode{MATCH v=dst(r) Where r.starttime<max(collect(vout IN out(v) | vout.endtime)) and r.starttime>=timestamp1 and r.endtime<=timestamp2}), restricting our search to the specific day when a particular attack occurred based on the attack descriptions provided by the datasets.

\begin{figure*}[t]
    \centering
    \begin{adjustbox}{width=\textwidth}
        \includegraphics[clip]{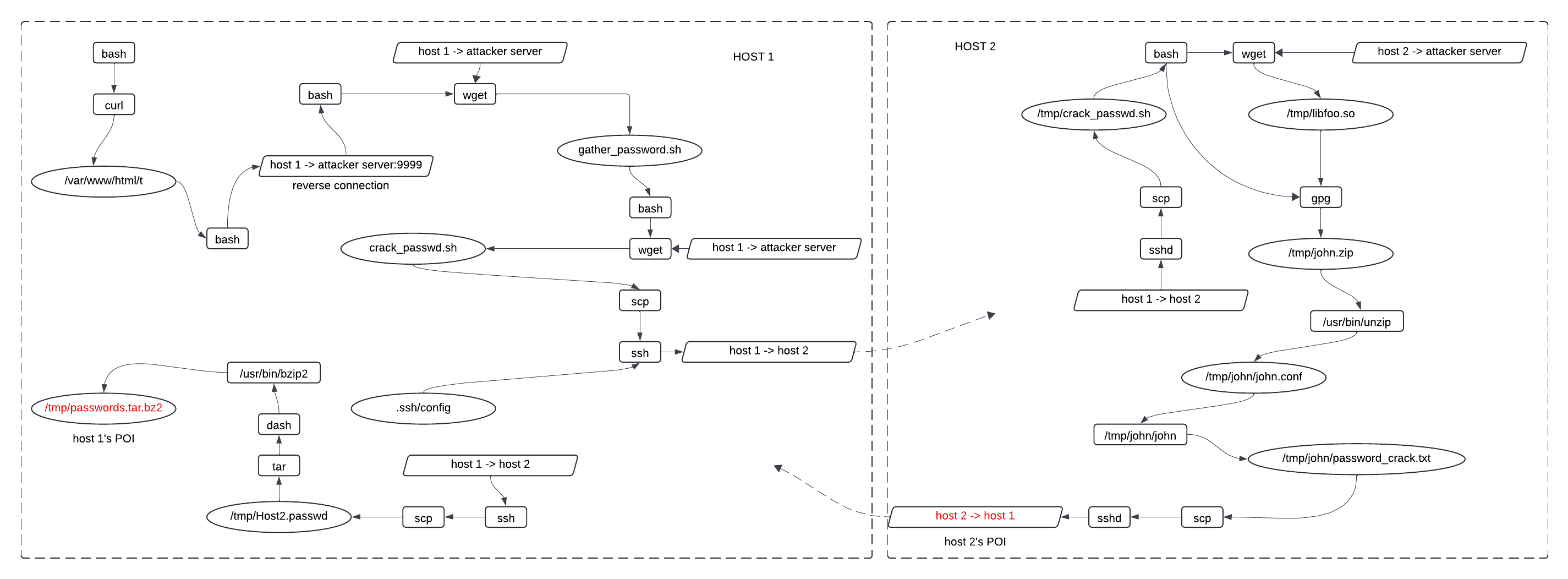}
    \end{adjustbox}
    \caption{\pg identified by \tool for Password Crack attack (non-critical edges are omitted)}
    \label{fig:Password Crack}
\end{figure*}

\myparatight{Case Study}
\rev{We use password crack attack as a case study, \cref{fig:Password Crack} shows critical edges of it. We illustrate, step-by-step, how ProGQL queries can be formulated in an interactive query style to perform provenance analysis for attack investigation.}

\begin{itemize}[noitemsep, topsep=1pt, partopsep=1pt, listparindent=\parindent, leftmargin=*]

\item \textit{Step 1 - Backward BFS from the POI}:
Starting from the point of interest (POI), \incode{/tmp/passwords.tar.bz2} on host1, ProGQL performs a backward BFS with temporal constraints:
\begin{lstlisting}[caption={},captionpos=b,label={query:step1}]
MATCH (p:Process)-[st:FileEvent{id:15035}]->(f:File{name:"/tmp/passwords.tar.bz2", hostid:"1"})
        BFS (r IN backward(f) | MATCH v=dst(r) WHERE r.starttime<max(collect(vout IN out(v) | vout.endtime))) YIELD g1
        RETURN g1
\end{lstlisting}
This query produces a \href{https://github.com/ProGQL/ProGQL/blob/main/technical%20report/Password%20Crack/neo4jpwd-back1.svg}{\pg} with 154 vertices and 3117 edges.

\item \textit{Step 2 - Edge weighting and impact score propagation}:
We next assign weights and propagate impact scores across the backward graph:
\begin{lstlisting}[caption={},captionpos=b,label={query:step2}]
MATCH (p:Process)-[st:FileEvent{id:15035}]->(f:File{name:"/tmp/passwords.tar.bz2", hostid:"1"})
        BFS (r IN backward(f) | MATCH v=dst(r) WHERE r.starttime<max(collect(vout IN out(v) | vout.endtime))) YIELD g1
        UNWIND g1 AS e
        SET e.weight=projection(1/(abs(r.amount-st.amount)+0.0001),ln(1+1/abs(r.endtime-st.endtime)),count(out(v))/count(in(v)))
        MATCH u=src(e) SET u.rel=reduce(sum = 0, o IN out(u) | sum+o.weight*dst(o).rel)
        RETURN g1
\end{lstlisting}
This step highlights critical dependencies by assigning high scores to influential nodes in \href{https://github.com/ProGQL/ProGQL/blob/main/technical%20report/Password%20Crack/neo4jpwd-back-weight1.svg}{\pg}.

\item \textit{Step 3 - Backward + forward analysis with entry selection}:
\tool then selects the top-ranking candidate entry nodes and performs forward BFS from them. Finally, it intersects the backward and forward graphs to isolate the most relevant \pg:
\begin{lstlisting}[caption={},captionpos=b,label={query:step3}]
MATCH (p:Process)-[st:FileEvent{id:15035}]->(f:File{name:"/tmp/passwords.tar.bz2", hostid:"1"})
        BFS (r IN backward(f) | MATCH v=dst(r) WHERE r.starttime<max(collect(vout IN out(v) | vout.endtime))) YIELD g1
        UNWIND g1 AS e
        SET e.weight=projection(1/(abs(r.amount-st.amount)+0.0001),ln(1+1/abs(r.endtime-st.endtime)),count(out(v))/count(in(v)))
        MATCH u=src(e) SET u.rel=reduce(sum = 0, o IN out(u) | sum+o.weight*dst(o).rel)
        RETURN g1
intersect
WITH entry = (MATCH n in nodes(r) WHERE count(in(n))=0 ORDER BY n.rel DESC LIMIT 15)
        BFS (re IN forward(entry) | MATCH u=src(re) WHERE re.endtime>min(collect(uin IN in(u) | uin.starttime)) and re.starttime<1724731846719889370) yield g2
        RETURN g2
\end{lstlisting}
The resulting \href{https://github.com/ProGQL/ProGQL/blob/main/technical%20report/Password%20Crack/neo4jpwd-vm1.svg}{\pg} has 39 vertices and 86 edges.

\item \textit{Step 4 - Multi-host correlation}: 
Repeating the same analysis on host2 and unioning the graphs from host1 and host2 yields a concise multi-host attack \pg.
\begin{lstlisting}[caption={},captionpos=b,label={query:step4}]
MATCH (p:Process)-[st:FileEvent{id:15035}]->(f:File{name:"/tmp/passwords.tar.bz2", hostid:"1"})
        BFS (r IN backward(f) | MATCH v=dst(r) WHERE r.starttime<max(collect(vout IN out(v) | vout.endtime))) YIELD g1
        UNWIND g1 AS e
        SET e.weight=projection(1/(abs(r.amount-st.amount)+0.0001),ln(1+1/abs(r.endtime-st.endtime)),count(out(v))/count(in(v)))
        MATCH u=src(e) SET u.rel=reduce(sum = 0, o IN out(u) | sum+o.weight*dst(o).rel)
        RETURN g1
intersect
WITH entry = (MATCH n in nodes(r) WHERE count(in(n))=0 ORDER BY n.rel DESC LIMIT 15)
        BFS (re IN forward(entry) | MATCH u=src(re) WHERE re.endtime>min(collect(uin IN in(u) | uin.starttime)) and re.starttime<1724731846719889370) yield g2
        RETURN g2
UNION
(MATCH (p:Process)-[st:NetworkEvent{id:100005}]->(f:Network{srcip:"192.168.1.128/32",dstip:"192.168.1.131/32",hostid:"2"})
        BFS (r IN backward(f) | MATCH v=dst(r) WHERE r.starttime<max(collect(vout IN out(v) | vout.endtime))) YIELD g1
        UNWIND g1 AS e
        SET e.weight=projection(1/(abs(r.amount-st.amount)+0.0001),ln(1+1/abs(r.endtime-st.endtime)),count(out(v))/count(in(v)))
        MATCH u=src(e) SET u.rel=reduce(sum = 0, o IN out(u) | sum+o.weight*dst(o).rel)
        RETURN g1
intersect
WITH entry = (MATCH n in nodes(r) WHERE count(in(n))=0 ORDER BY n.rel DESC LIMIT 15)
        BFS (re IN forward(entry) | MATCH u=src(re) WHERE re.endtime>min(collect(uin IN in(u) | uin.starttime)) and re.starttime<1724731846712161377) yield g2
        RETURN g2)

\end{lstlisting}
The union \href{https://github.com/ProGQL/ProGQL/blob/main/technical%20report/Password%20Crack/neo4jpwd-vmunion.svg}{\pg} has 124 vertices and 282 edges.
\end{itemize}

\subsection{RQ2: Comparison with Cyher}
\label{subsec:rq2}
We choose to compare against Cypher because other domain-specific languages either share its level of expressiveness (e.g., AIQL~\cite{aiql}) or lack graph search capabilities altogether (e.g., SAQL~\cite{saql}). To evaluate language effectiveness, we conduct a series of experiments that assess both expressiveness and the ability to accurately reconstruct attack sequences. Our results show that only \tool can generate concise graphs that effectively capture attack patterns from audit logs. 
Cypher, despite its flexibility, requires overly verbose queries and fails to match \tool’s precision and performance.
We next discuss the complexity of Cypher's query construction and its execution performance. 

\subsubsection{Complexity of Query Construction}
\tool is specifically designed to streamline the process of querying audit logs for attack detection by efficiently performing constrained graph search, computing edge weights, and propagating values through the graph. In contrast, Cypher lacks the expressiveness for these critical functions, making it incapable of identifying attack entry points without extensive manual intervention. To compensate for this limitation in our evaluations, we manually identified the top ranked nodes and constructed corresponding Cypher queries to assess their ability to generate a concise graph that reveals the attack sequences. Unlike \tool, Cypher necessitates the construction of multiple, complex queries to achieve the same goal. For example, rewriting \tool Query~\ref{query:pwd} in Cypher requires 32 separate Cypher queries. Specifically, the forward BFS in \tool Query~\ref{query:pwd} (Line 8) needs to be repeated 15 times using queries like Query~\cite{progqlforwardcypher}. This not only increases the complexity of the query construction but also makes the process more error-prone and less maintainable.

\begin{table}[!t]
	\centering
	\caption{\pgs produced by \tool and \depimpact}
 \label{tab:comparesize}
	\resizebox{\linewidth}{!}
 {
        \begin{tabular}{l|rrr|rrr}
        \hline
 \multicolumn{1}{c|}{\multirow{2}{*}{\textbf{Attack} }} &  \multicolumn{3}{c|}{\textbf{\depimpact}}  & \multicolumn{3}{|c}{\textbf{\tool}}     \\ \cline{2-4} \cline{5-7}
&\multicolumn{1}{c}{\textbf{FP}} & \multicolumn{1}{c}{\textbf{FN}}   & \multicolumn{1}{c}{\textbf{Edges}} & \multicolumn{1}{|c}{\textbf{FP}} & \multicolumn{1}{c}{\textbf{FN}}   & \multicolumn{1}{c}{\textbf{Edges}}\\ \hline
Wget Executable              &    56        &  0  &68  & 53& 0 &65  \\     
        Illegal Storage            &   206  &   0 & 212 & 206 &0 &212         \\
        Hide File  & 141  & 0  &149  & 38 & 0 & 46  \\
        Steal Information  &  328 &  0  & 335 &43 & 0 &50   \\
        Backdoor Download  & 128  &  0  & 134 &41 &0 &47   \\
        Annoying Server User  &  254 &  0  & 266 &34 &0 &46 \\
        Password Crack  &  257 &  1  & 293 & 245 &0 & 282  \\
        Data Leakage  & 268  &  0  & 285 &264 &0 & 281  \\
        Vpn Filter  & 575  &  4  & 581 & 627 &0 & 637  \\
        Theia Case 1  &12,734  & 0   &12,741  & 332 & 0 &339    \\
        Theia Case 3  & 797,509  &   0 &797,513 &2 & 0& 6   \\
        Fivedirections Case 1  &  119,861 & 0   &119,865  & 10 & 0 &14    \\
        Fivedirections Case 3  &173,996 & 0   &173,998  &15 &0 &17   \\
        Trace Case 5  & \multicolumn{3}{c|}{\text{OOM}}   & 471 & 0 &474 \\
        \hline
        \textbf{Average}  & 85,101 &  0  & 85,111 & 170 & 0 & 180 \\
        \hline
        \end{tabular}
	}
\end{table}

\begin{table}[!t]
	\centering
	\caption{Statistics of the execution time (second)}
 \label{tab:comparetime}
	\resizebox{0.8\linewidth}{!}{%
 \tiny{
        \begin{tabular}{lrr}
        \hline
        \multicolumn{1}{c}{\textbf{Attack}}     & \multicolumn{1}{c}{\textbf{\depimpact}} & \multicolumn{1}{c}{\textbf{\tool}}  \\ 
        \hline
        Wget Executable              & 232           & 24     \\     
        Illegal Storage            & 237    & 50            \\
        Hide File  & 349  & 36     \\
        Steal Information  & 268  & 55     \\
        Backdoor Download  & 317  & 22    \\
        Annoying Server User  & 631  & 629      \\
        Password Crack  &  9 & 23     \\
        Data Leakage  & 9  & 16      \\
        Vpn Filter  &  10 &  23     \\
        Theia Case 1  &2,995  & 155     \\
        Theia Case 3  & 2,906 & 320     \\
        Fivedirections Case 1  & 242  & 8    \\
        Fivedirections Case 3  & 245  & 210     \\
        Trace Case 5  & OOM  & 1,069  \\ \hline

       \textbf{Average} & 650  & 189 \\ \hline 
        \end{tabular}
        }%
}
\end{table}

\begin{table}[!t]
	\centering
	\caption{Results of RQ1 variants with various weight fitering conditions}
        \label{tab:edgefilter}
        \resizebox{\linewidth}{!}{%
	\small
        \begin{tabular}{lrrrrr}
        \hline
        \multicolumn{1}{c}{\textbf{Attack}}     & \multicolumn{1}{c}{\textbf{CPR (\# E) }} & \multicolumn{1}{c}{\textbf{\tool$_w$ (\# E)}}  & \multicolumn{1}{c}{\textbf{\# CritEdges}} &  \multicolumn{1}{c}{\textbf{\# CritKept}} & \multicolumn{1}{c}{\textbf{\% CritLost}}\\ \midrule
        Wget Executable              & 1,967 &167 &12         & 6   & 50\%  \\     
        Illegal Storage            & 880    & 153   & 6 & 4 &33\%          \\
        Hide File  & 1,420  & 173   & 8 &6 &25\%   \\
        Steal Information  & 1,650  & 168  & 7 &4 &43\%   \\
        Backdoor Download  & 1,773  & 175   & 6 &4 &33\%   \\
        Annoying Server User  & 6,901  & 762   & 12 &9 &25\%   \\
        Password Crack  & 995  & 281   & 37 &22 &41\%   \\
        Data Leakage  &1,676 &234 & 17 &8 &53\%  \\
        Vpn Filter  & 4,897 & 664 & 10 & 3 & 70\% \\
        Theia Case 1  & 27,397  & 6,525   & 7 &4 &43\%    \\
        Theia Case 3  & 8  & 6   & 4 &4 &0\%    \\
        Fivedirections Case 1  & 319  & 244   & 4 &2 &50\%    \\
        Fivedirections Case 3  & 17  & 2   & 2 &0 &100\%    \\
        Trace Case 5  & 827  & 41   & 3 &2 &33\% \\ \midrule 
        \textbf{Average} & 3,623  & 685 & 10 & 6 & 43\% \\ \hline
        \end{tabular}
        }%
\end{table}

\subsubsection{Inefficiency and Inaccuracy in Cypher}

\myparatight{False Positives}
\cref{tab:comparesizecypher} shows Cypher generates a significant larger number of false positives in the output graphs (21,481 for Password Crack, 21,539 for Data Leakage, and 24,827 for VPN Filter). 
\rev{Cypher (even with APOC) is fundamentally path-oriented: it enumerates complete paths from the start node. Filters like \incode{WHERE incoming.starttime < updatedMaxEndTime} are evaluated after a path has been generated, based only on the nodes and relationships in that path. This means Cypher cannot enforce recursive, edge-dependent constraints during the traversal itself. Instead, it allows traversal to continue along edges that should already have been pruned, and only later removes them from the returned results. As a consequence, Cypher produces additional edges (false positives) because it lacks \tool’s ability to prune dynamically at each recursion step. In contrast, \tool is subgraph-oriented: its BFS operator applies edge-dependent constraints at every step, ensuring that only qualified edges are included.}

\myparatight{Case Study - Controlled depth comparison with Cypher}
\rev{
To further highlight \tool’s precision, we conducted a controlled depth comparison by running backward BFS from the POI on host1 with a maximum depth of nine. We then compared the provenance graphs produced by \tool and by a best-effort Cypher approximation of the same query.
Below we show the Cypher query we constructed as a best-effort approximation of the \tool backward BFS; this highlights the structural similarities but also makes evident where Cypher fails to enforce recursive, edge-dependent constraints.}

At depths (1--8), the difference is not immediately visible. At depth 9, the provenance graph returned by \tool has 119 vertices and 2,998 edges \protect\href{https://github.com/ProGQL/ProGQL/blob/main/technical%20report/Password%20Crack/neo4jpwd-back1-depth-9.svg}{\pg}, while the \protect\href{https://github.com/ProGQL/ProGQL/blob/main/technical%20report/Password%20Crack/CypherPwdByDepth-9.svg}{\pg} returned by Cypher has 138 vertices and 6600 edges.
At shallow depths (1--8), the difference is not immediately visible, because most paths near the POI happen to overlap with causally valid edges, so Cypher’s outputs look superficially similar. However, starting at depth 9, the divergence becomes pronounced. Cypher continues expanding along many irrelevant paths - such as PAM libraries, NSS lookups, and background socket activity - because it evaluates constraints only after paths are generated, rather than pruning edges recursively during traversal. \tool, by contrast, consistently excludes these false positives, maintaining a concise subgraph that reflects only attack-relevant dependencies.

\begin{lstlisting}[caption={},captionpos=b,label={query:step1}]
MATCH (p:Process)-[st:FileEvent{optype:"write"}]->(start:File{name:"/tmp/passwords.tar.bz2", hostid:"1"}) 

// Carry the POI and initialize a max end-time constant.This matches the same starting condition as in ProGQL.
WITH start, 1724731846719889370 AS initialMaxEndTime 

// APOC expands all backward paths up to maxLevel. Here, YIELD path always means the entire path from start to the current node, not just the most recent hop. This whole-path semantics is where Cypher diverges from ProGQL. ProGQL expands step by step with constraints applied at each hop; APOC dumps a superset of all traversed edges into path.

CALL apoc.path.expandConfig(start, { 
    relationshipFilter: "<", 
    minLevel: 0,  
    maxLevel: 9, 
    bfs: true,  
    uniqueness: "NODE_GLOBAL", 
    filterStartNode: false 
}) YIELD path  

// Label the path (visitedPath) and extract its current endpoint (currentNode).
WITH start, path AS visitedPath, last(nodes(path)) AS currentNode, initialMaxEndTime 

// Intended to prune traversal depth.
WHERE length(visitedPath) < 9
// Tries to restrict expansion to edges that are in the current path. But because visitedPath is the whole path, this condition still admits edges from earlier steps - including ones that would have been pruned if filtering happened inline (like in ProGQL). This is the main FP source.
OPTIONAL MATCH (currentNode)-[outgoing]->() 
WHERE outgoing IN relationships(visitedPath) OR currentNode = start

// Aggregate outgoing edges to compute the temporal bound. But since outgoing edges include superset ones, the max can be inflated.
WITH start, currentNode, outgoing, max(outgoing.endtime) AS maxEndtime, initialMaxEndTime, visitedPath 

// Set the temporal constraint. At the POI, use the initial cutoff; otherwise, inherit the max endtime.
WITH start, currentNode, outgoing, initialMaxEndTime, visitedPath,
     CASE 
	 WHEN currentNode = start THEN initialMaxEndTime
	 ELSE COALESCE(maxEndtime, initialMaxEndTime) 
     END AS updatedMaxEndTime 
     
// Apply causal pruning (only accept edges that end before the cutoff). But pruning happens after expansion, so invalid edges were already collected into visitedPath.
MATCH (currentNode)<-[incoming]-() 
WHERE incoming.starttime < updatedMaxEndTime

// Gather pruned incoming edges and project source/sink sets. This simulates causal filtering, but again false positives that slipped into visitedPath remain in play.
WITH start, currentNode, COLLECT(incoming) AS updatedVisitedRels, updatedMaxEndTime, visitedPath 
WITH start, currentNode, updatedVisitedRels, updatedMaxEndTime, 
     [r IN updatedVisitedRels | startNode(r)] AS sourceNodes, 
     [r IN updatedVisitedRels | endNode(r)] AS sinkNodes, visitedPath 

// Output the traversal frontier.
RETURN DISTINCT currentNode, updatedVisitedRels AS visitedRels, updatedMaxEndTime, sourceNodes, sinkNodes;

\end{lstlisting}

\begin{figure*}[htbp]
    \centering
    
    \begin{subfigure}{0.32\textwidth}
        \includegraphics[width=\linewidth]{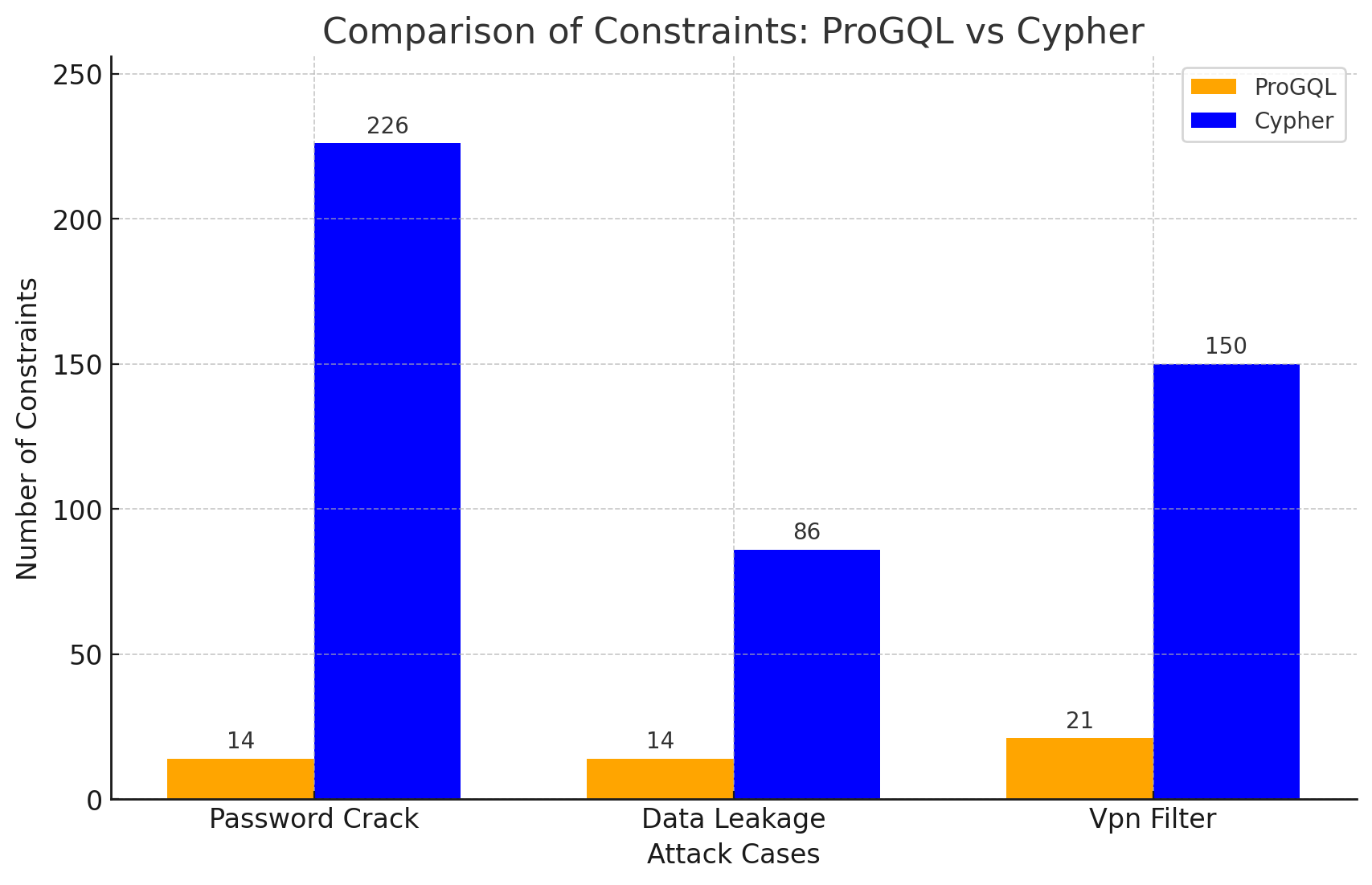}
        \caption{Comparison of Constraints}
    \end{subfigure}
    \hfill
    \begin{subfigure}{0.32\textwidth}
        \includegraphics[width=\linewidth]{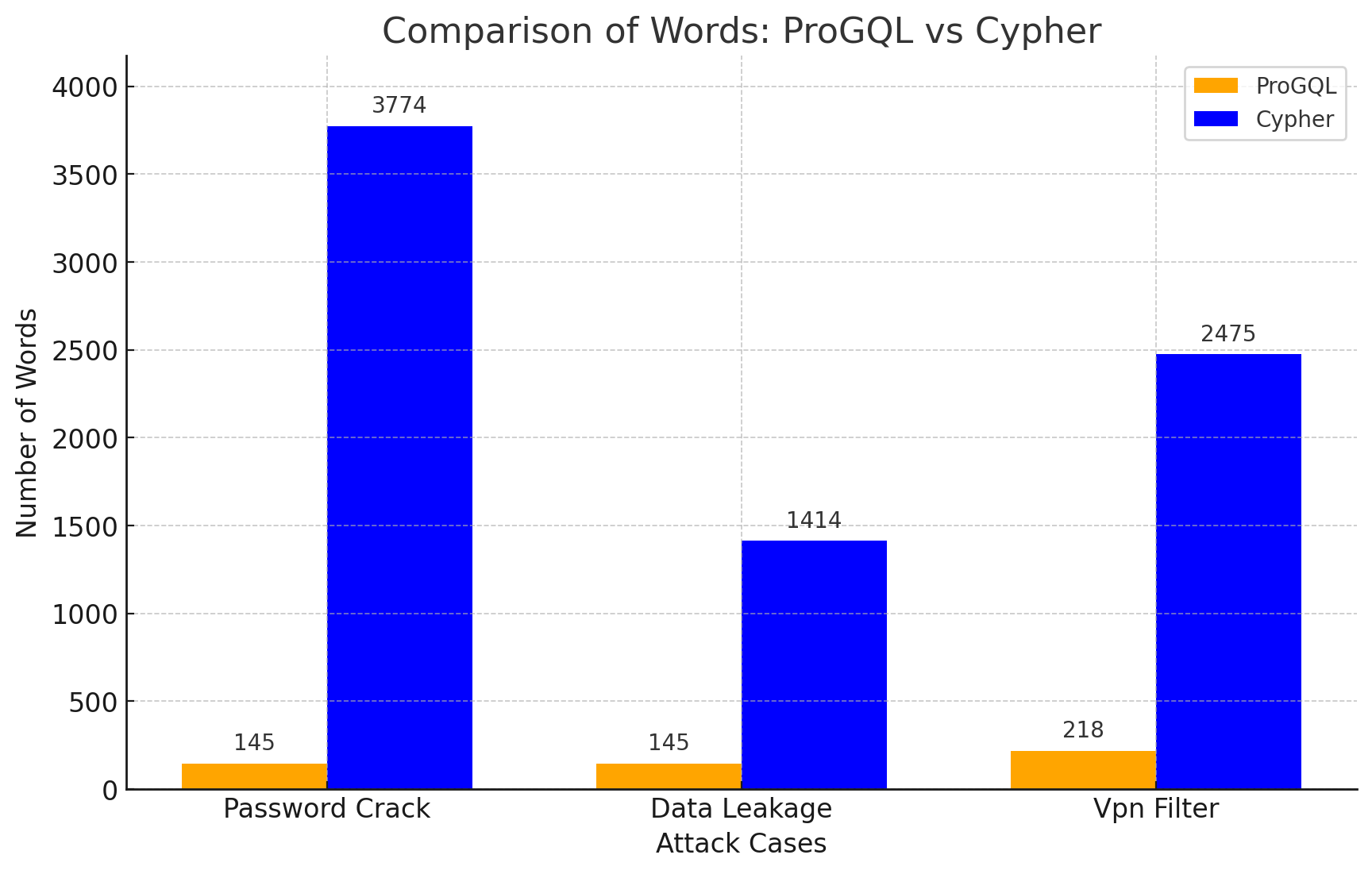}
        \caption{Comparison of Words}
    \end{subfigure}
    \hfill
    \begin{subfigure}{0.32\textwidth}
        \includegraphics[width=\linewidth]{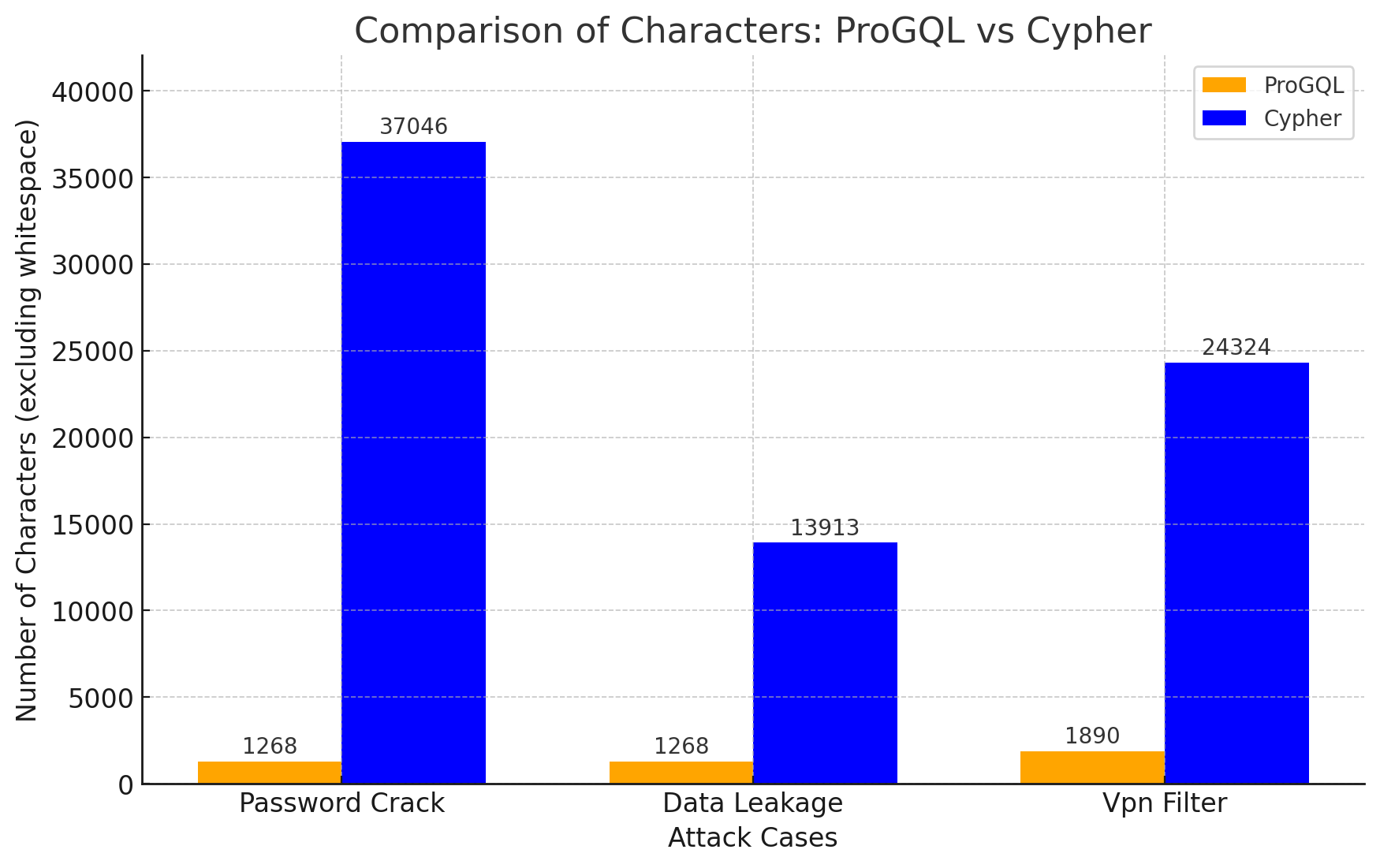}
        \caption{Comparison of Characters}
    \end{subfigure}
    
    \caption{\centering Conciseness evaluation of queries written in \tool and Cypher}
    \label{fig:conciseness}
\end{figure*}

\noindent\rev{\myparatight{Conciseness Evaluation}
We evaluated query conciseness using three metrics: the number of constraints, words, and characters. A constraint is defined as any atomic restriction that filters, bounds, or enforces semantics in a query, including \incode{WHERE} clauses (temporal, structural, or equality conditions), typed edge/node restrictions (e.g., 
\incode{FileEvent{optype:"write"}}), aggregations for pruning (e.g., \incode{max(...)}, \incode{min(...)}), and bounds such as \incode{LIMIT} or depth cutoffs. Words are counted as tokens separated by whitespace, and characters as any non-whitespace symbols. As shown in \cref{fig:conciseness}, across three multi-host attack cases, \tool consistently outperforms Cypher in brevity, requiring on average $9\times$ fewer constraints, $15\times$ fewer words, and $17\times$ fewer characters.
}

\myparatight{Execution Time and Memory Consumption}
The experiments reveal a stark contrast in performance between the two languages. \tool not only executes significantly faster (e.g., 21s vs. 478s on average) but also consumes less memory (1.77GB vs. 2.96GB on average). For other larger datasets used in the experiments, Cypher’s performance deteriorates severely, as it was unable to complete the queries within a reasonable timeframe (over 2 hours), making it impractical for attack investigation in large-scale environments.

\subsection{RQ3: Comparison with SOTA \pa Technique}
\label{subsec:rq3}

To demonstrate the effectiveness of \tool in revealing the attack sequence, we compare \tool with the state-of-the-art (SOTA) technique: \depimpact\cite{depimpact}. For each attack, \tool ranks the nodes based on their impact scores and chooses the entry nodes in each of the three system entity categories to perform forward \pa analysis from the nodes in the order of decreasing impact scores. 
\tool stops choosing a new node once all the critical events that represent attack steps have been identified in the \pg or if the newly chosen node causes the output \pg to include significantly more edges. 
We apply the same entry nodes picking logic to \depimpact to make a fair comparison.

\subsubsection{Effectiveness Comparison}
\cref{tab:comparesize} shows the comparison of the \pgs produced by \tool and \depimpact. 
The results show that although both techniques do not miss any critical edge that represents attack steps, the \pgs produced by \tool are much smaller on the DARPA datasets. 
This is because DARPA attack cases span multiple days (\eg the Theia dataset contains logs for 8 days), \tool can target the attacks accurately with less processing time by adding temporal constraints or system library constraints in the \incode{Where} clause, which can filter out more irrelevant edges. 
On the other side, \depimpact produced $473\times$ larger graphs on average since it does not have the flexibility to add more filters. 
Moreover, \depimpact encountered \textbf{out-of-memory errors (OOM)} when running on the DARPA datasets even if we increased the heap size to 100GB.
This shows the major limitation of \depimpact that loads all the event data into memory.

\subsubsection{Efficiency Comparison}
To evaluate the efficiency of \tool, we conducted experiments
for each case on each database, repeating each test three
times and calculating the average execution time. 
We picked the optimal performances among the six types of database backends (See \cref{subsec:evalsetup}) and compared them with \depimpact.
To make the evaluation fair, we disabled/cleared database caches for \tool after finishing running each \tool query, and we excluded logs loading and parsing time from \depimpact.
The results are shown in \cref{tab:comparetime}.  
Overall, \tool executes faster than \depimpact (189s vs. 650s).
These results indicate that \textbf{\textit{even though \tool executes search through queries with databases, the index of the edges and the incremental graph search make up for the loss of performing search in the memory}}.
On the other side, \depimpact spent much more time on the DarpaTC datasets because it lacks the flexibility to add the temporal filter.

\subsubsection{Memory Consumption Comparison}
\cref{tab:comparememory} compares the memory consumption of \depimpact with \tool with all the six types of database backends.
We can observe that the memory consumption of \depimpact is ~$8\times$ greater than \tool, which shows the superiority of \tool that utilizes a database backend.
Note that building a large memory pool is usually impractical for many companies due to the much higher costs.
We can also observe that all databases require much less memories compared to \depimpact.

\begin{table*}[!t]
	\centering
	\caption{Statistic of memory consumption (GB)}
 \label{tab:comparememory}
	\resizebox{0.8\linewidth}{!}{%
 \tiny{
        \begin{tabular}{l|r|rrrrr}
        \hline
\multirow{2}{*}{\textbf{Attack}} &  \multirow{2}{*}{\textbf{\depimpact}}  & \multicolumn{5}{c}{\textbf{\tool}}     \\ \cline{3-7}
&&\multicolumn{1}{c}{\textbf{PostgreSQL}} & \multicolumn{1}{c}{\textbf{MyRocks}}   & \multicolumn{1}{c}{\textbf{Mariadb}} & \multicolumn{1}{c}{\textbf{Neo4j}} & \multicolumn{1}{c}{\textbf{Nebula}}  \\ \hline
        Wget Executable &66.26   & 0.82   & 0.84 &0.81 &1.61 & 0.84  \\
        Illegal Storage &71.08   & 0.97   & 0.98 &0.94 &1.94 & 0.95  \\
        Hide File & 74.76  & 0.89  & 0.93 & 0.97 &1.63 & 0.93 \\
        Steal Information   &70.70   & 1.05  & 1.43 & 1.03 &2.01 &1.01 \\        
        Backdoor Download  & 72.19  & 0.79  & 0.89 &0.82 &1.44 &0.87 \\
        Annoying Server User    & 79.48  &15.14 & 15.38 & 13.32 &15.07 & 19.82 \\
        
        Password Crack & 1.67 & 0.80  & 0.80  & 0.89 & 1.94 &  0.82 \\
        Data Leakage  & 1.61  & 0.88  & 0.91 & 0.83 & 1.45 & 0.79  \\
        Vpn Filter  & 2.25  & 0.86  & 1.34 & 0.85 & 1.91 & 0.86  \\
        
        Theia Case 1  &26.17   & 2.72   & 1.78 & 2.72 &3.00 &  1.54 \\
        Theia Case 3    & 24.71  & 4.38  & 3.49 & NA &8.30 & 3.26  \\
        Fivedirections Case 1   & 5.98  &0.36   & 0.36 &0.35  &1.46 & 0.30 \\
        Fivedirections Case 3   & 7.21  &3.25   & 1.64 & 1.75 &5.32 & 1.71 \\
        Trace Case 5    & OOM  & 7.92  & 6.95 & 7.48 &24.76 & NA  \\
        \hline
        \textbf{Average} &  38.77 & 2.92 & 2.69 & 2.52 & 5.13 & 2.59\\
        \hline
        \end{tabular}
        }%
	}
\end{table*}

\subsection{RQ4: \rev{Flexibility in Edge Weight Computation}}

\label{subsec:rq4}
To demonstrate the flexibility of \tool, we implement variants of the \pa used in RQ1 by adopting different weight-based mechanisms to filter out irrelevant edges.
Specifically, we compose a variant for CPR~\cite{reduction} (without edge filtering based on weights) and a variant (\tool$_w$) that performs backward \pa and retains edges with weights greater than or equal to 0.5 (\autoref{query:weight}).
This shows that by easily changing the filtering condition specified in the \emph{$\langle$Where$\rangle$} rule, \tool can easily re-implement any existing weight-based \pa~\cite{reduction,hassan2019nodoze,liu2018priotracker}.

\begin{lstlisting}[caption={\tool Query with weight filtering},captionpos=b,label={query:weight}]
MATCH (p:Process)-[st:FileEvent{optype:"write"}]->(f:File{name:"/home/fs/sysrep_random"})
	BFS (r IN backward(f) | MATCH v=dst(r) WHERE r.starttime<max(collect(vout IN out(v) | vout.endtime))) YIELD g1
	UNWIND g1 AS e
	SET e.weight=projection(1/(abs(r.amount-st.amount)+0.0001),ln(1+1/abs(r.endtime-st.endtime)),count(out(v))/count(in(v)))
	WITH e WHERE e.weight >=0.5
	RETURN g1
\end{lstlisting}

\autoref{tab:edgefilter} shows the results for the two variants of \tool. 
Column ``CPR (\# E)'' shows the number of edges after applying CPR's edge merge (Section~\ref{subsec:engine}).
Column ``\tool (\# E)'' shows the number of edges output by \tool$_w$.
Column ``\# CritEdges'' displays the critical edges that represent attack steps of the \pgs, while Column ``\# CritKept'' shows how many of these critical edges are preserved by \tool$_w$. 
Finally, Column ``\% CritLost'' quantifies the percentage of critical edges missed by \tool$_w$. 

As we can see, although edge merging reduces parallel edges between nodes within a given time threshold, the resulting graph still contains a significant number of irrelevant edges (averaging 3,623) compared to the ground truth, which typically includes only 10 edges.
\tool$_w$ greatly improves over CPR by removing edges using computed weights (from an average of 3,623 to 685), but it also causes significant loss of critical edges. 
Notably, some scenarios, such as ``Fivedirections Case 3'' experienced a complete loss of critical edges (100\%). 
These results emphasize the need to carefully choosing weight computations for edge filering, and demonstrate the flexibility of \tool in expressing \pa techniques with different weight computations.

\myparatight{Supporting Different Edge Weights with Variable $K$ in Top-K Entry Node Selection}
\rev{As shown in a recent study~\cite{depimpact}, the choice of edge-weight assignments and the selection of $K$ in top-$K$ entry node selection are critical for preserving attack-related edges while filtering out irrelevant ones.
Thus, we further conduct experiments on the choices of edge weights combined with different $K$ in top-$K$ entry node selection across three multi-host attack cases: Password Crack, Data Leakage, and Vpn Filter.
\tool can naturally express both feature-based edge weighting and entry node candidate selection. For example, top-$K$ entry node selection can be directly expressed using \incode{LIMIT} with proper sorting, making it straightforward to select the most likely attack entry points. Similarly, feature selection can be declaratively controlled via projection functions in the query.
For example, 
\incode{SET e.weight = projection(1/(abs(r.amount - st.amount) + 0.0001))}
uses only the data flow feature (amount) for weight computation, rather than the 3-feature expressed in Query~\ref{query:pwd}.}

\begin{figure*}[htbp]
    \centering   
    \begin{subfigure}{0.48\textwidth}
        \includegraphics[width=\linewidth]{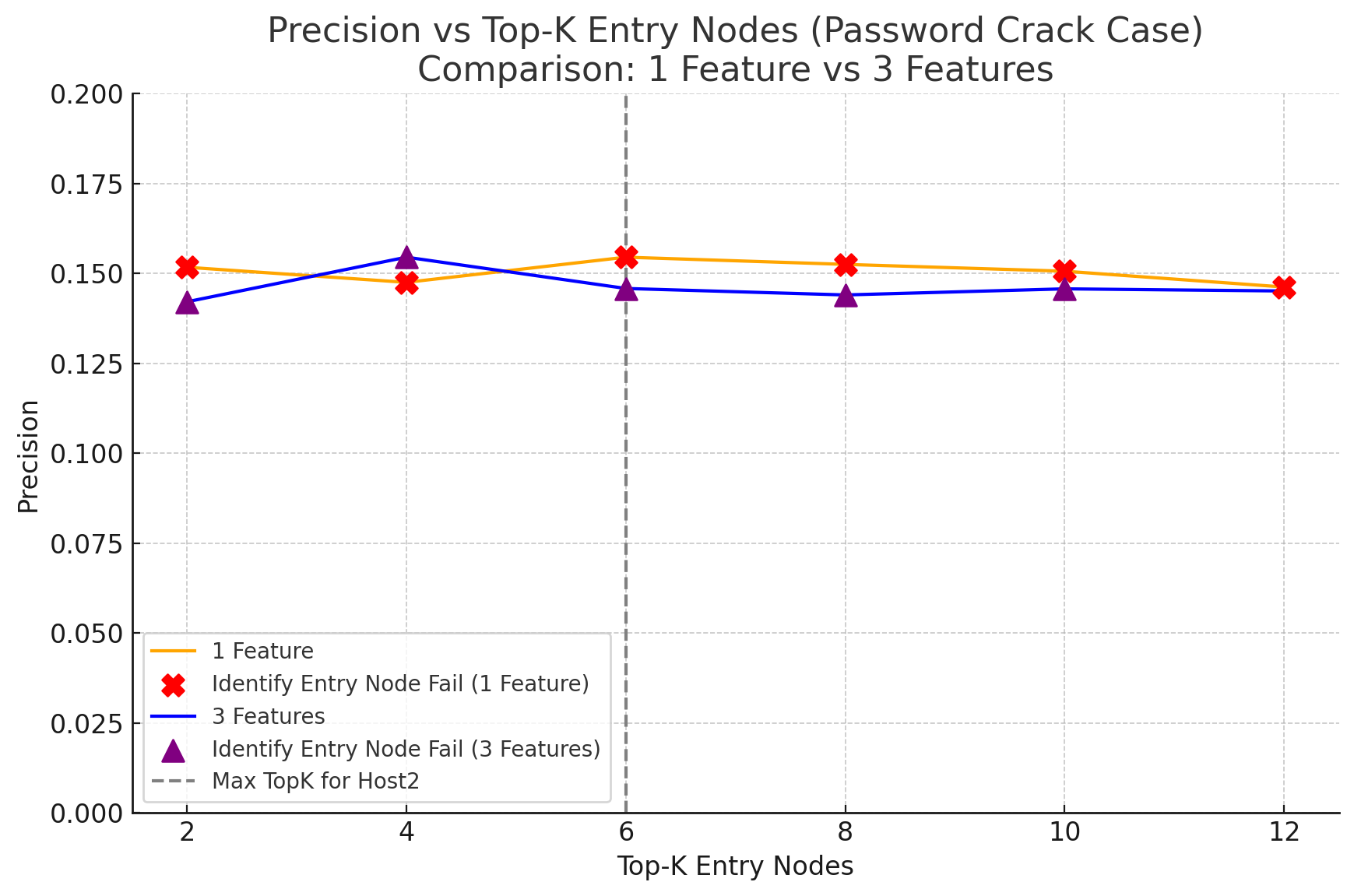}
        \caption{Precision vs Top-K (Password Crack)}
    \end{subfigure}%
    \hfill
    \begin{subfigure}{0.48\textwidth}
        \includegraphics[width=\linewidth]{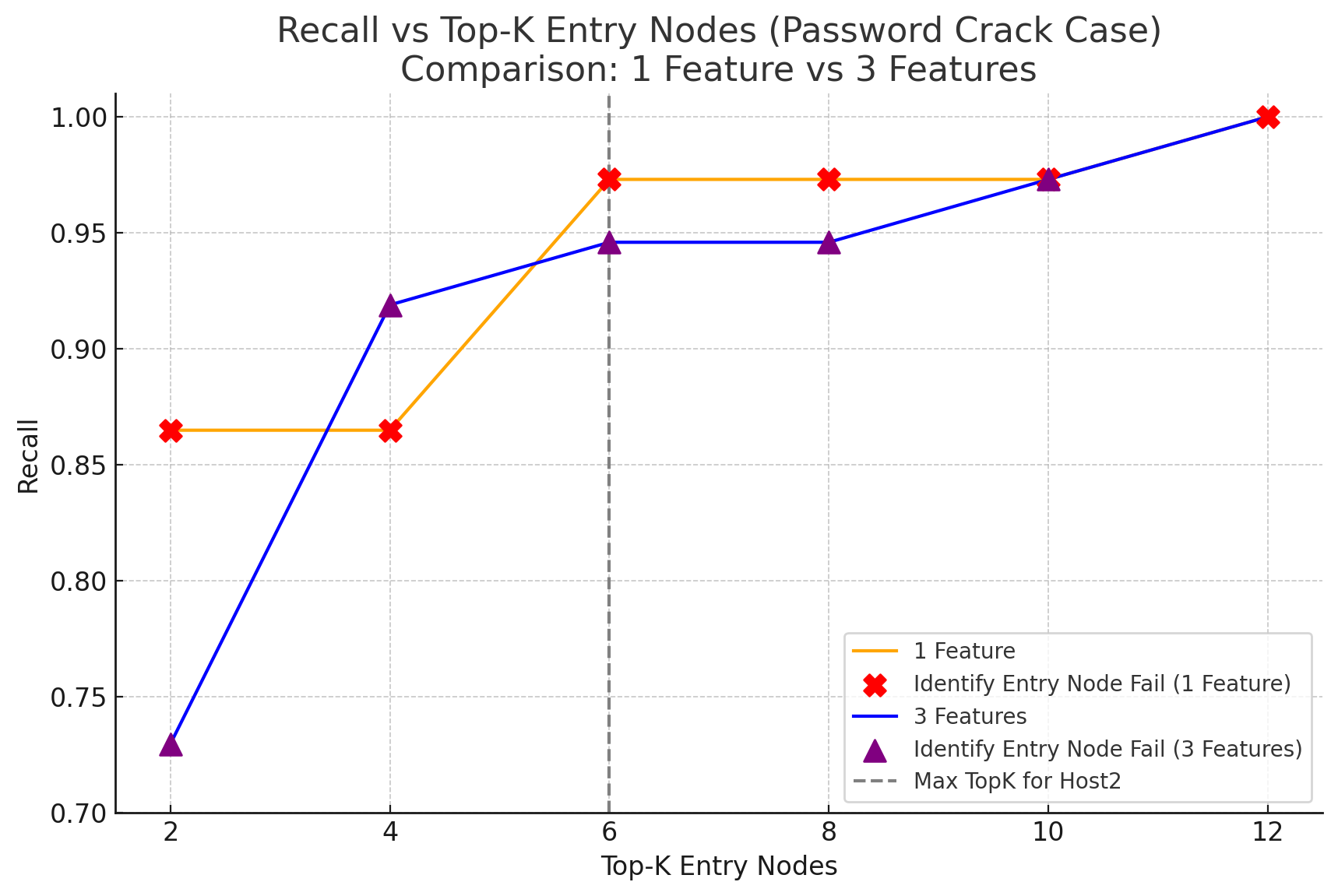}
        \caption{Recall vs Top-K (Password Crack)}
    \end{subfigure}
    
    \begin{subfigure}{0.48\textwidth}
        \includegraphics[width=\linewidth]{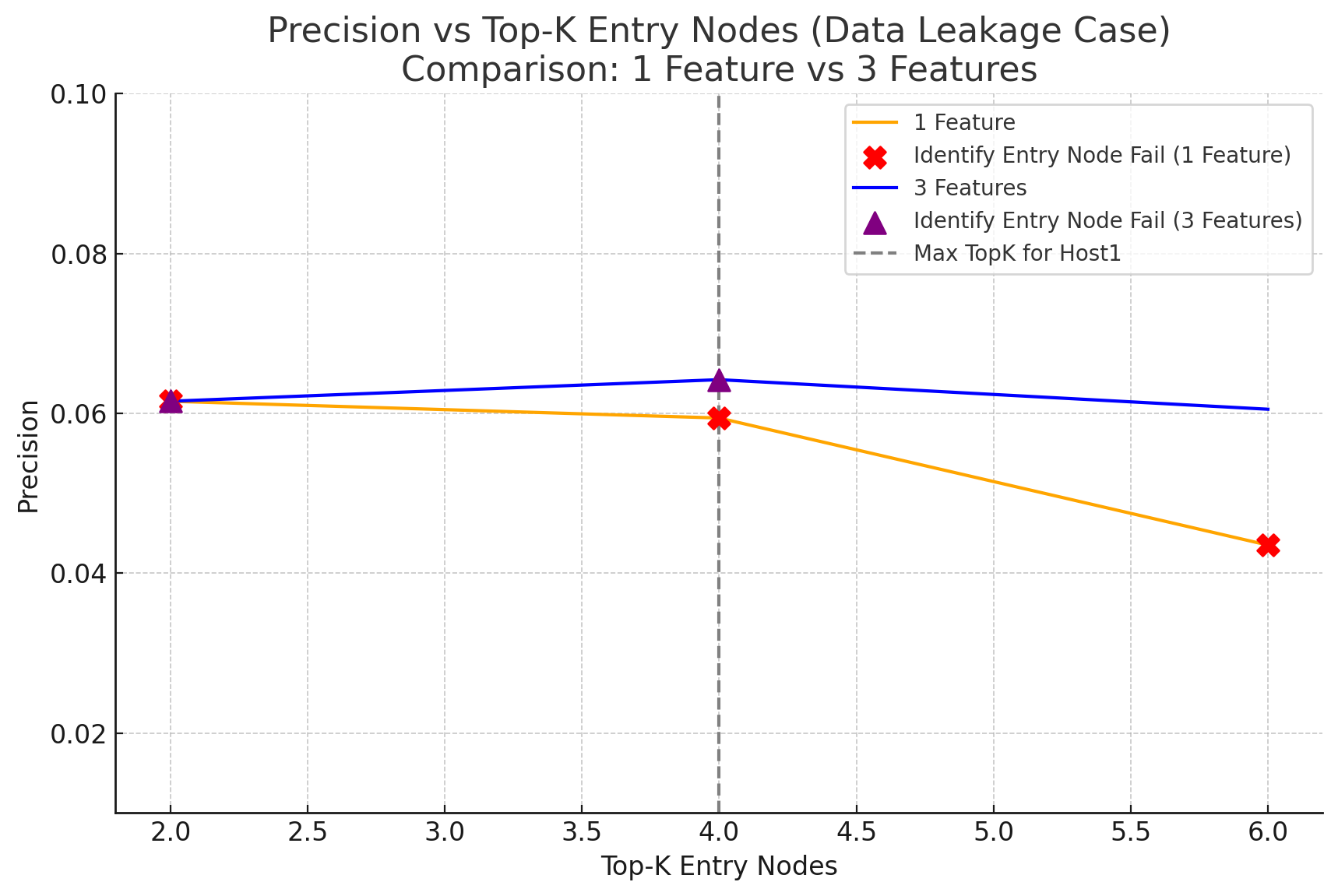}
        \caption{Precision vs Top-K (Data Leakage)}
    \end{subfigure}%
    \hfill
    \begin{subfigure}{0.48\textwidth}
        \includegraphics[width=\linewidth]{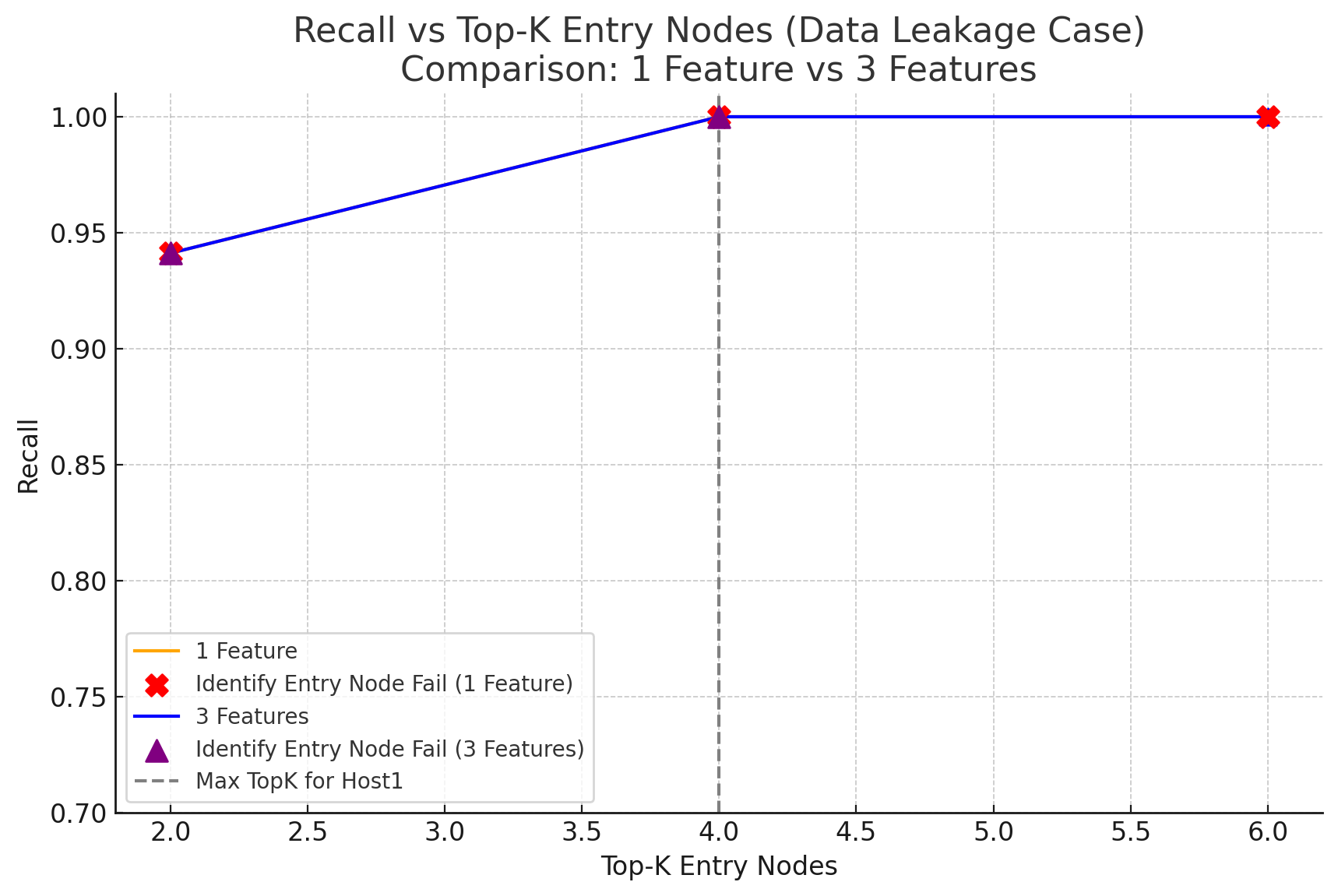}
        \caption{Recall vs Top-K (Data Leakage)}
    \end{subfigure}
    
    \begin{subfigure}{0.48\textwidth}
        \includegraphics[width=\linewidth]{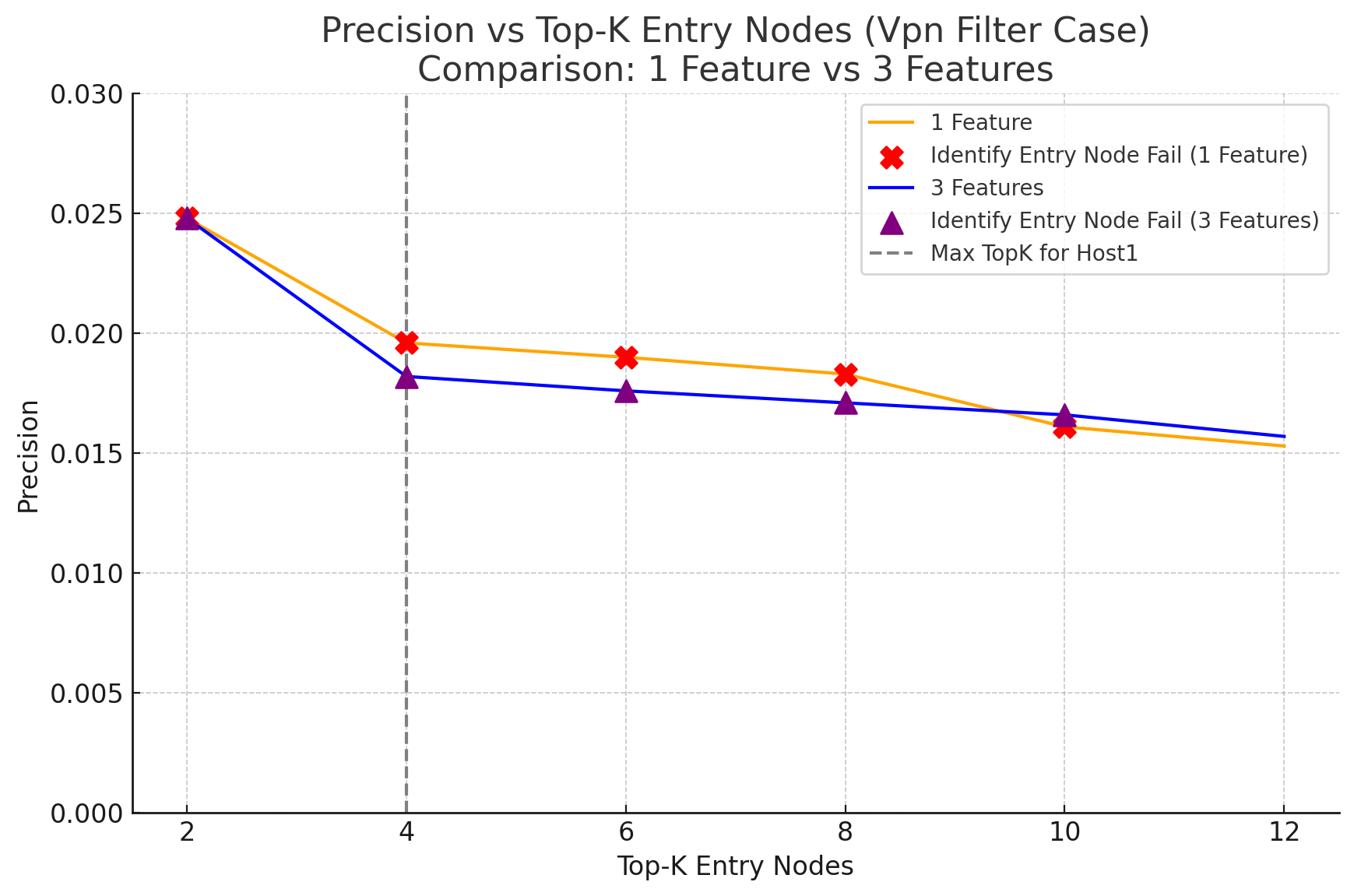}
        \caption{Precision vs Top-K (VPN Filter)}
    \end{subfigure}%
    \hfill
    \begin{subfigure}{0.48\textwidth}
        \includegraphics[width=\linewidth]{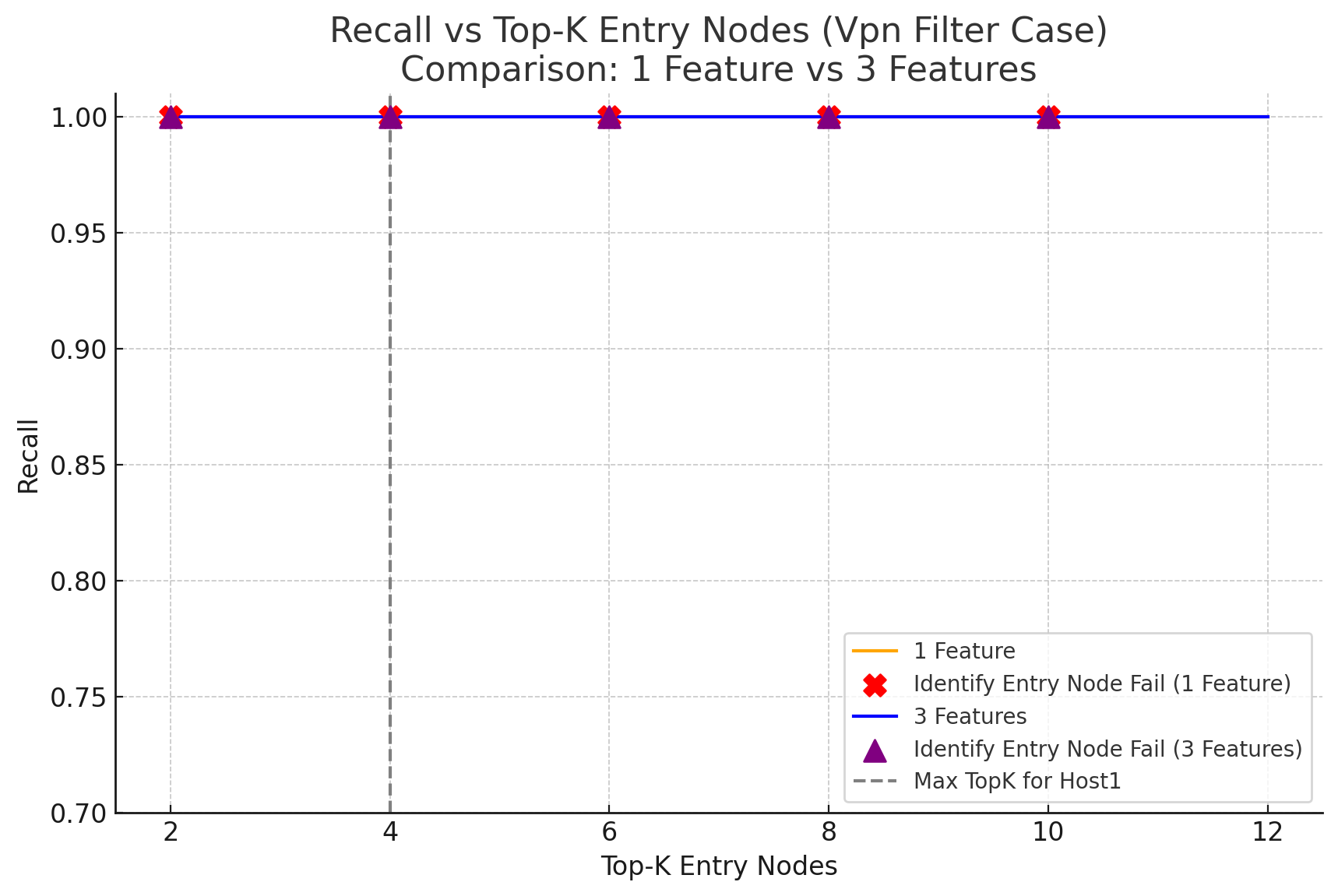}
        \caption{Recall vs Top-K (VPN Filter)}
    \end{subfigure}
    
    \caption{\centering Feature and TopK entry nodes ablation analysis across three attack cases (Password Crack, Data Leakage, VPN Filter)}
    \label{fig:precisionrecall}
\end{figure*}

\rev{As shown in \cref{fig:precisionrecall}, we observed a natural trade-off between recall and precision in top-$K$ entry node selection: a larger $K$ improves recall but also introduces more false positives, since additional non-critical edges are included. The benefit of 3-feature weighting is evident in this trade-off. While both 1-feature and 3-feature weight computations eventually achieve high recalls, the 3-feature weight computation often reaches comparable recall earlier and maintains consistently higher precision by suppressing spurious edges. Moreover, 1-feature weight computation fails to identify ground-truth entry nodes for Password Crack and Data Leakage, whereas 3-feature weight computation avoids such failures, demonstrating greater robustness. Together, these results show that combining temporal, structural, and data-flow features yields more efficient prioritization, higher precision, and more reliable recall than relying on a single feature.}

\begin{table*}[!t]
	\centering
	\caption{Statistics of database execution time (second)}
 \label{tab:comparedb}
	\resizebox{0.8\linewidth}{!}{%
        \fontsize{4}{4}\selectfont
        \begin{tabular}{lrrrrr}
        \hline
        \multicolumn{1}{c}{\textbf{Attack}}     & \multicolumn{1}{c}{\textbf{PostgreSQL}} & \multicolumn{1}{c}{\textbf{MyRocks}} &
        \multicolumn{1}{c}{\textbf{Mariadb} } 
        &\multicolumn{1}{c}{\textbf{Neo4j}} &
        \multicolumn{1}{c}{\textbf{Nebula}} \\ \hline
        Wget Executable     &60     &  26    &  24     & 99    &236    \\     
        Illegal Storage       &82     & 59    &  50    &  128    &226          \\
        Hide File  & 57  & 36  &   36 &  90   & 155    \\
        Steal Information  &94  & 57    & 55    & 140   &394 \\
        Backdoor Download  &57  & 22  &  22   &93    &88   \\
        Annoying Server User  &680  & 810   & 734     & 629   & 3,533   \\
        Password Crack  &80   &134   & 122 & 23 &  564  \\
        Data Leakage  & 27  & 41  & 35 & 16 & 151   \\
        Vpn Filter  & 72  & 116  & 112 & 23 & 469   \\
        Theia Case 1  &155 & 230  &1,370   &242  & 180   \\
        Theia Case 3  &491  &  684  &  $>$2 hr    &  320   & 4,821   \\
        Fivedirections Case 1  & 26  &13  &8 &56 &   29    \\
        Fivedirections Case 3  & 222  &435 &365 &210 &1,358  \\
        Trace Case 5  &1,271  &2,823 & 2,396& 1,069& $>$2 hr  \\ \hline 
        \textbf{Average} & 241 & 392 & 410 & 224 & 939\\ \hline
        \end{tabular}
        
	}
\end{table*}

\subsection{RQ5: Database Backend Comparison }
\label{subsec:rq5}

We compare the performance of \tool using 5 types of database backends: PostgreSQL, MyRocks, MariaDB, Neo4j, and Nebula.
The comparison of these databases is based on measurements of memory consumption (\cref{tab:comparememory}) and running performance (\cref{tab:comparedb}). 
Since some database backends may run for a long time for certain queries, we terminated the query execution if no results are obtained within two hours.

\myparatight{Runtime Performance Comparison}
Based on \cref{tab:comparedb}, it becomes evident that among the databases completing all experiments within two hours, Neo4j demonstrates superior running performance (with an average of 224 seconds) over other databases, particularly when traversing through a substantial volume of nodes/edges. 
This performance superiority can be attributed to Neo4j's optimized graph traversal mechanism: Neo4j traversal API for graph search, which offers a flexible and expressive approach to define graph traversal logic. 
A interesting observation is that in scenarios where extensive volume is not a prerequisite, MyRocks and Mariadb surpass other databases in running performance.
It's interesting to observe that when Nebula is tasked with traversing a substantial volume of nodes/edges, it tends to persist indefinitely with relatively low memory consumption. For instance, it consistently consumes about 24GB even after running for 3 hours, and the mechanism behind this prolonged runtime may be caused by the employed memory consumption cap. 
In contrast, Neo4j automatically allocates more memory to ensure expedited traversals.

\myparatight{Memory Consumption Comparison}
According to \cref{tab:comparememory}, it is evident that among the databases where all experiments are successfully executed within two hours, MyRocks exhibits the lowest memory usage (average of 2.69GB), while Neo4j demonstrates the highest memory consumption (average of 5.13GB). 
The memory consumption for the executions that are longer than 2 hours are marked as ``NA''. 
This discrepancy can be attributed to the utilization of the Neo4j traversal API for graph search.
Nevertheless, in comparison to other databases, Neo4j's higher memory usage can be linked to additional memory requirements for indexing, aiming at optimizing the graph traversal speed. 
The variance in memory requirements is also influenced by the unique indexing structures employed by relational databases for similar operations.

\myparatight{Recommended Database Backends}
While no database is the clear winner for both memory consumption and execution time, we recommend using either Neo4j or PostgresSQL.
In scenarios where memory consumption is not an issue (up to 25GB), Neo4j’s fast search offers superior efficiency, benefiting from today’s cheaper memory.
Otherwise, PostgreSQL provides slightly slower search performance but operates within a smaller memory footprint (up to 15GB).


\section{Discussion}
\label{sec:discussion}


\noindent\textbf{Data Compression}.
In our implementation, we employ some existing data compression techniques such as CPR~\cite{reduction}.
There are several more recent works~\cite{reduction2,reduction3,reduction4,depcomm} that achieve even better data compression rates.
These compression approaches preserve the dependencies used by \pa, and thus will not affect our search results and can be directly integrated with our \tool framework. 


\noindent\textbf{Mixed Database Backend}.
Our evaluation shows that relational databases are faster in searching for specific events while graph databases are more efficient in edge traversal tasks.
One potential optimization is to combine the strengths of these two types of databases: using relational databases' indexes to find specific starting nodes and then using graph databases for graph traversal. 
Although Neo4j supports third-party indexes, the lack of native integration leads to additional overhead. Native index support in graph databases remains an area in need of further research.

\rev{
\noindent\textbf{Extensibility for Dependency-Centric Graph Domains}.
Beyond APT attack investigation, \tool's operators naturally generalize to any dependency-centric graph domain. For instance, constrained recursive search has also been applied to knowledge graphs or RDF database to discover multi-hop relationships under semantic or temporal constraints. Similarly, feature-driven edge weighting generalizes to applications in program analysis, supply-chain dependency tracking, or data lineage systems, where analysts must flexibly choose relevant edge features (degree, timestamps, amounts, etc.) to emphasize in traversal. \tool’s weight computation operator is designed to support this flexibility: it allows arbitrary combinations of neighbor-based features, encompassing prior designs such as the 3-feature and the 1-feature weight computations described in Section~\ref{subsec:rq4}.}

\noindent\textbf{Parallelism for Further Scaling}.
Our current implementation considers relational databases and graph databases as our database backend. 
As shown in the recent studies~\cite{aiql,aiqldemo,prosql,aptrace}, parallelism, which is orthogonal to what we propose in this paper, can further speed up the query search if the auditing event data is stored based on its domain-specific properties such as hosts and time. 
Furthermore, if the \pgs built from different sub-queries in a \tool query has no data dependence on each other, which requires program analysis on the \tool query, then we can leverage ``map-reduce'' strategy to execute multiple search in parallel and then performs graph merge in the end. 
We also would like to explore the direction on adopting Hadoop~\cite{ghazi2015hadoop} and MapReduce~\cite{dean2008mapreduce} to further scale up the data storage and speed up query search.

\section{Related Work}
\label{sec:literature}

\noindent\textbf{Domain-Specific Query Language for Security Applications}.
There exist domain-specific languages in a variety of security fields that have a well-established corpus of low level
algorithms, such as threat descriptions~\cite{cybox,taxii,stix}, 
secure overlay networks~\cite{mace,networklang}, and network intrusions~\cite{chimera,lambda,Sommer:2014:HAE:2663716.2663735,Vallentin:2016:VUP:2930611.2930634}.
These languages provide specialized constructs for their particular problem domain.
Recent works~\cite{aiql,aiqldemo,saql,saqldemo,aptrace,prosql} also provide domain-specific languages to express various patterns to detect attack behaviors from system audit logs.
In contrast to these languages, 
the novelty of \tool focuses on provenance analysis by providing specialized constructs for recursive graph search and value propagation,
which existing graph query languages~\cite{graphquery,graphquery2,cypher} also do not support.

\noindent\textbf{Provenance Analysis}.
King et al.~\cite{backtracking,backtracking2} proposed a backward causality analysis technique to perform intrusion analysis by automatically reconstructing a series of events that are dependent on a user-specified POI event.
Following this research, recent efforts have been made to mitigate the dependency explosion problem~\cite{beep,ma2016protracer,mcitracking,ji2017rain,ji2018enabling,liu2018priotracker,hassan2019nodoze}. 
In Section~\ref{sec:eval}, we have shown that \tool can express complex algorithm like \depimpact~\cite{depimpact}.
Our proposed \tool system can well support different \pa algorithms by expressing their search constraints except for intrusive system modifications like binary instrumentation~\cite{mcitracking,ma2016protracer}, and can also seamlessly integrate their optimizations by consuming the optimized events produced by these techniques.

\noindent\textbf{Database Query Language}.
Database query languages are designed for general-purpose data search. 
Relational databases based on SQL and SPARQL~\cite{postgresql,sql,sql-tuning,sparcle05} provide language constructs for joins, facilitating specification of relationships among activities.
Graph databases such as Neo4j~\cite{neo4j} provide language constructs in their query language Cypher~\cite{cypher} for finding paths or nodes in graphs.
NoSQL tools such as DynamoDB~\cite{dynamodb} and MongoDB~\cite{chodorow2013mongodb} provide simpler language for fast data fetches based on keys. 
There are also other query languages for spatio-temporal databases~\cite{spatiotemporaldb,tquel,timesql}.
None of these languages can express customized graph search and support value propagation like \tool, which is critical for \pa. Based on prior languages (e.g., Cypher), CQL and SQL/PGQ~\cite{deutsch2022graph} are proposed towards a standard query language on property graphs. Our proposed language structs built on top of Cypher comply with the standard and can be easily integrated.

\noindent\textbf{System Analysis Language}.
Besides academia, industry has recently released several query languages designed for fine-grained system analysis.
OSQuery~\cite{osquery,osquerysec} lets analysts use SQL queries to probe the real-time system status.
Elasticsearch~\cite{elasticsearch} and Splunk~\cite{splunk} are log-analysis platforms that index general application logs,
and provide a keyword-based search language to perform data search.
Similar to database query languages, these languages lack of the capability to express \pa.

\section{Conclusion}
We propose \tool, a framework that supports customizable \pa on system audit logs and improves the scalability of \pa through incremental graph search. 
Specifically, the \tool framework provides a domain-specific language that includes novel language constructs for constrained graph search, edge weight computation, value propagation, and graph merge. 
To scale up the search over a colossal amount of system audit logs, the \tool framework imports the logs into a database backend formed by either relational databases or graph databases, and provides a query engine to achieve incremental graph search with the help of the database backend.
Our evaluations show that our \tool language provides better expressiveness than SOTA graph query languages such as Cypher in expressing a diverse set of attack behaviors, 
and the comparison with the state-of-the-art \pa demonstrates the \tool framework's significant scalability improvement ($8$ times saving of memory consumption without penalty on runtime performance).

\bibliographystyle{IEEEtran}
\bibliography{./bibliography/refs}


\end{document}